%% file: templateArxiv.tex
\documentclass{article}

\usepackage{PRIMEarxiv}

\usepackage[utf8]{inputenc} 
\usepackage[T1]{fontenc}    
\usepackage{booktabs}       
\usepackage{amsfonts}       
\usepackage{nicefrac}       
\usepackage{microtype}      
\usepackage{lipsum}
\usepackage{fancyhdr}       
\usepackage{graphicx}       
\graphicspath{{media/}}     
\usepackage{ltablex}
\usepackage{caption}
\usepackage{subcaption}
\usepackage{xcolor}
\usepackage{multirow}
\usepackage{lscape} 
\usepackage{arydshln}
\usepackage{hyperref}       
\usepackage[font=small,labelfont=bf]{caption}
\usepackage{authblk}

\title{Synthetic data generation for a longitudinal cohort study -- Evaluation, method extension and reproduction of published data analysis results}

\author[1,2]{Lisa Kühnel}
\author[1]{Julian Schneider}
\author[3]{Ines Perrar}
\author[4]{Tim Adams}
\author[5]{Fabian Prasser}
\author[3]{Ute Nöthlings}
\author[4,6]{Holger Fröhlich}
\author[1,7]{Juliane Fluck}
\affil[1]{Knowledge Management, ZB MED -- Information Centre for Life Sciences, 50931 Cologne,
Germany; kuehnel@zbmed.de, schneider@zbmed.de, fluck@zbmed.de}
\affil[2]{Graduate School DILS, Bielefeld Institute for Bioinformatics Infrastructure (BIBI), Faculty of Technology, Bielefeld University, 33615 Bielefeld, Germany}
\affil[3]{Institute of Nutritional and Food Sciences -- Nutritional Epidemiology, University of Bonn, 53115 Bonn, Germany; iperrar@uni-bonn.de, u.noethlings@uni-bonn.de}
\affil[4]{Department of Bioinformatics, Fraunhofer Institute for Algorithms and Scientific Computing SCAI, 53757 Sankt Augustin, Germany; tim.adams@scai.fraunhofer.de, holger.fröhlich@scai.fraunhofer.de}
\affil[5]{Medical Informatics Group, Berlin Institute of Health at Charité – Universitätsmedizin Berlin, 10117 Berlin, Germany; fabian.prasser@bih-charite.de}
\affil[6]{Bonn-Aachen International Center for IT, University of Bonn, Friedrich Hirzebruch-Allee 6, 53115 Bonn, Germany}
\affil[7]{The Agricultural Faculty, University of Bonn, 53115 Bonn, Germany}

\begin{document}
\maketitle

\begin{abstract}
Access to individual-level health data is essential for gaining new insights and advancing science. In particular, modern methods based on artificial intelligence rely on the availability of and access to large datasets. In the health sector, access to individual-level data is often challenging due to privacy concerns. A promising alternative is the generation of fully synthetic data, i.e. data generated through a randomised process that have similar statistical properties as the original data, but do not have a one-to-one correspondence with the original individual-level records. In this study, we use a state-of-the-art synthetic data generation method and perform in-depth quality analyses of the generated data for a specific use case in the field of nutrition. We demonstrate the need for careful analyses of synthetic data that go beyond descriptive statistics and provide valuable insights into how to realise the full potential of synthetic datasets. By extending the methods, but also by thoroughly analysing the effects of sampling from a trained model, we are able to largely reproduce significant real-world analysis results in the chosen use case.
\end{abstract}

\keywords{Synthetic Health Data \and Nutritional Studies \and Epidemiological Data \and Machine Learning}

\input{chapter/01_intro}
\input{chapter/02_methods}
\input{chapter/03_results}

\input{chapter/04_discussion}

\input{chapter/05_conclusion.tex}

\section*{Availability}
We provide the source code for the adapted algorithm under \url{https://github.com/nfdi4health/vambn-extensions-evaluations/}. 

\section*{Acknowledgements}
This work was done as part of the NFDI4Health Consortium (\url{www.nfdi4health.de}). We gratefully acknowledge the financial support of the Deutsche Forschungsgemeinschaft (DFG, German Research Foundation) –- project number 442326535. \newline
The DONALD Study is financially supported by the Ministry of Science and Research of North Rhine-Westphalia, Germany. Trend analyses of the original data were part of a project funded by the German Federal Ministry of Food and Agriculture (BMEL) through the Federal Office for Agriculture and Food (BLE), grant 2816HS024.

\bibliographystyle{unsrt}  
\bibliography{references}  

\input{chapter/06_appendix.tex}

\end{document}

%% file: chapter/01_intro.tex
\section{Introduction}
In biomedical research, scientific progress is often limited by the availability and quality of data. The results of any study can only be as good as the data on which the statistical analysis was based. Additionally, machine learning methods - including deep learning - are completely dependent on the quality and amount of the available training data. For some application fields it is extremely hard to obtain enough data to be able to draw any conclusions, for example in the case of rare diseases. In particular, to realise the full potential of deep learning methods, the amount of data should be larger than for more traditional machine learning approaches~\cite{deep_learning_needs_data}. To increase availability of medical data, a legally compliant mechanism for sharing it between different institutions is needed. Furthermore, being able to share the data used in a publication is an important step in making its results reproducible in terms of the FAIR principles \cite{wilkinson_fair_2016}. However, the storage, processing, and sharing of individual-level health data is tightly regulated and restricted by law, as health-related data are generally considered to be highly sensitive (e.g.~\cite{eu16sensitivedata} Art. 9).

Sharing personal health data in compliance with laws and regulations, such as the European Union (EU) General Data Protection Regulation (GDPR), usually requires informed consent. However, this is often not feasible, for example if data is to be analysed retrospectively at a large scale. As an alternative, data can be anonymised in such a way that it cannot be traced back to specific individuals anymore. This typically requires significant modifications, e.g. by removing direct identifiers (such as names), and by coarsening indirect identifiers (such as age or geographic region). And this approach inevitably requires balancing the reduction of risks achieved through the removal of information with associated reductions in the utility of the data \cite{aggarwal_k-anonymity_2005}. This also means that a complete anonymisation might not always be possible to achieve for many types of data, with genetic information being a common example.

An alternative way to share data could be to use synthetic data generation methods that have been proven efficient in the past, e.g. by \cite{lei_mri-only_2019, wendland_generation_2022, sood_realistic_2020, goncalves_generation_2020, rankin_reliability_2020}. In this process, instead of modifying the original data to make them harder to re-identify, a completely new dataset is created, ideally with similar statistical properties as the real life data. 

In this study, we apply and adapt state-of-the-art algorithms to generate synthetic data for a defined use case, for which several downstream analyses have already been performed on the real data and their results have been published. 
In addition to analysing summary statistics of the dataset, this gives us the opportunity to gain deep insights into the possibilities and limitations of synthetic data in the specific use case. 

The original data that are used are parts of the data collected within the \textit{Dortmund nutritional and anthropometric longitudinally designed (DONALD) study}, which is an ongoing nutritional cohort study capturing information about the diet and health of children in Dortmund, Germany since 1985~\cite{buyken12donald}. Participants are recruited as newborns, and then accompanied until young adulthood to get a holistic picture of their developing health and the dietary factors that influence it. The subset used during this study includes dietary data with focus on sugar intake based on all dietary records of participants aged three through 18 years between 1985 and 2016. The resulting dataset consists of structured health data, where the same 33 variables have been recorded over fixed yearly intervals. 

The data collected by the DONALD study have already been used for several nutritional studies, such as the recent analysis of sugar intake time trends by Perrar \textit{et al.}~\cite{perrar_time_2019, perrar20donaldtrends}, which was performed on the same subset used for this work. Hence, this subset, that will from now on only be called DONALD data, has the following properties that need to be taken into account when searching an appropriate synthetic data generation method: It is longitudinal, because its data have been collected over a series of 16 visits. Additionally, it contains static variables that are only collected during the first visit. It is also heterogeneous, since its columns consist of various different data types. Finally, it is incomplete in terms of longitudinality, as not all participants attended each annual visit.

In the literature, a variety of different machine learning-based methods have been proposed to generate synthetic data, and we will discuss the following three classes that are commonly used (or combinations of them): probabilistic models, variational autoencoders, and generative adversarial networks (GANs). Especially GANs, originally developed by Goodfellow \textit{et al.}~\cite{goodfellow14gan}, are currently used in a variety of generative tasks and have proven successful in the generation of realistic fake images and in producing natural text data (e.g. \cite{liu_generative_2021,karras_analyzing_2020, ren_generating_2020, subramanian_adversarial_2017}). However, GANs have been proposed for the generation of continuous data \cite{goodfellow14gan}. Choi \textit{et al.} \cite{choi17medgan} address this by combining a GAN with an autoencoder in their architecture, which has been pre-trained to compress and reconstruct the original data. The resulting model, MedGAN, learns to generate EHRs. It provides an attractive solution to the generation of convincing electronic health records with continuous, binary and count features. Unfortunately, it cannot be used for the generation of DONALD data, since it is not able to generate longitudinal data. 

An alternative for longitudinal data has been developed by Esteban \textit{et al.}, who propose two different models (RGAN and RCGAN), both based on recurrent neural networks, to produce real-valued time series data \cite{esteban_real-valued_2017}. Even though the model is able to learn longitudinality, it cannot cope with heterogeneous data types, as well as a mixture of static and longitudinal variables that are typical in clinical studies. A further alternative is timeGAN, another GAN-based method that learns an embedding to represent longitudinal data in a lower dimensional space \cite{yoon2019time}. However, timeGAN is unable to generate static covariates and has not been applied to the medical domain yet -- hence, data incompleteness would potentially cause problems that require further work. 

Another promising method is called Variational Autoencoder Modular Bayesian Network (VAMBN), which has been, indeed, designed to generate fully synthetic data for mixed static and longitudinal datasets containing heterogeneous features with missing values \cite{gootjes20vambn}. To achieve this, a Heterogeneous-Incomplete Variational Autoencoder (HI-VAE) is combined with a Bayesian Network (BN). This works by splitting the data into modules, training a HI-VAE for each of them to produce encodings, and fitting the BN over all encodings of the modules. Longitudinality is handled by assigning the data of different visits to different modules. Due to its capabilities, VAMBN serves as a baseline algorithm for the present study. 

Because of the generative nature of the task -- in contrast to discriminative tasks -- the evaluation of the quality and usefulness of the data is not trivial, especially because it is performed on complex, heterogeneous health data. Different measures are used in different studies, probably due to different features having large variations in type and importance, which are highly dependent on the use case. Additionally, there is no standard terminology used by related studies. In this paper, we will follow the terminology from Georges-Filteau and Cirillo, who classify the metrics broadly into quantitative and qualitative metrics \cite{georges-filteau_synthetic_2020}. Here, qualitative analysis is based on visual inspection of the results done by field experts. For example, in \textit{preference judgement}, given a pair of two data points -- one real and one synthetic -- the aim is to choose the most realistic one \cite{choi17medgan}. A similar method of this category is called \textit{discrimination task}, where the expert is shown one data point at a time and needs to decide whether it is realistic or not. However, according to Borji, qualitative methods that are based on visual inspections are weak indicators and quantitative measures offer more convincing proofs of data quality at the dataset or sample level~\cite{borji_pros_2018}. \newline
Georges-Filteau and Cirillo~\cite{georges-filteau_synthetic_2020} further classify three subcategories of quantitative measures: comparisons between real and synthetic data on dataset level, comparisons of individual feature distributions, and utility metrics, which indicate whether the synthetic data can be used for real world analyses that were planned or already performed on the real dataset. 

Assessing the privacy of the data may be an even more difficult task, and is also handled differently in related works. While in some studies the risk of re-identification is assessed or controlled using empirical analyses (e.g. \cite{choi17medgan, goncalves20synthprivacy}), another established method known as \textit{differential privacy} \cite{dwork_calibrating_2006} can provide theoretical probabilistic bounds for various types of privacy risks. This method has been adapted to the area of deep learning by adding a specified amount of noise during training \cite{abadi_deep_2016}. Differential privacy can be integrated into all algorithms presented in this study. While acknowledging the importance of a more careful analysis of the privacy risks associated with synthetic data, we would like to point out that this is in practice a rather complex task, which we see out of the scope of this paper.  

In the present study, we make the following contributions: We apply the state-of-the-art algorithm VAMBN to generate synthetic data based on the DONALD dataset -- a dataset from a nutritional cohort study. We extend VAMBN with a long short-term memory (LSTM) layer \cite{hochreiter97lstm} to more effectively encode longitudinal parts of the data and show a significant increase in the ability to reproduce direct dependencies across time points. We evaluate our generated synthetic data on four different levels and show that while descriptive summary statistics and individual variable distributions can be efficiently reproduced with all chosen methods, direct dependencies can only be reproduced by our proposed extension. With this, we apply real-world experiments together with domain experts and gain valuable insights on how to exploit the potential of fully synthetic datasets. 

%% file: chapter/02_methods.tex
\section{Material and Methods}
\subsection{DONALD Data}
The method of the used DONALD dataset has been described in detail by Perrar \textit{et al.} \cite{perrar_time_2019, perrar20donaldtrends}. The main content of the data for each record is the nutrient intake, e.g. fat or carbohydrate intake, as well as the intake of different types of sugars, measured as a percentage of the total daily energy intake in kilocalories per day (\%E). 

\begin{figure}
    \centering
   \begin{subfigure}[b]{0.48\textwidth}
         \includegraphics[width=0.9\textwidth]{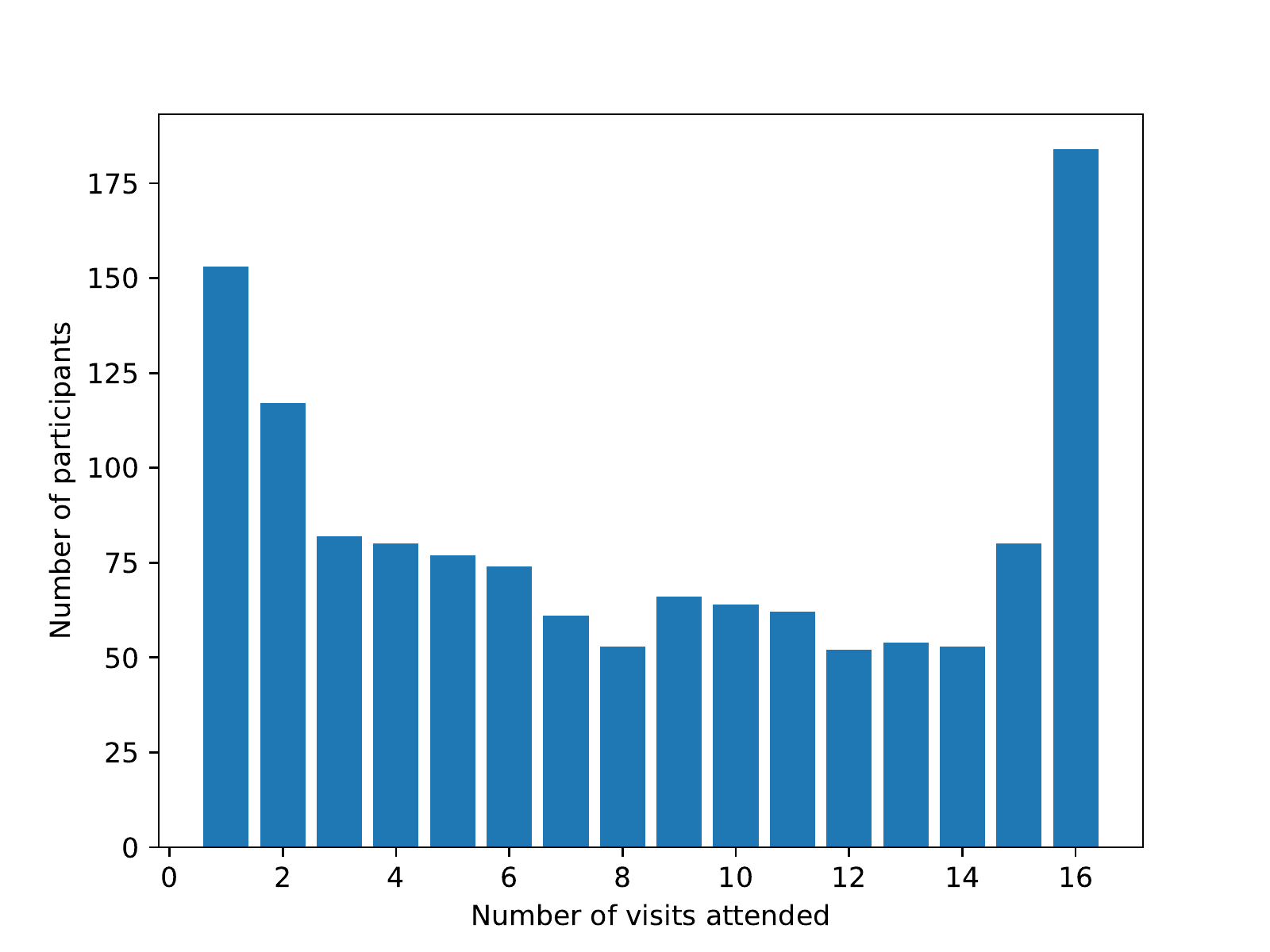}
         \caption{Total amount of attended visits}
         \label{fig:donald_missing_count}
     \end{subfigure}
     \begin{subfigure}[b]{0.48\textwidth}
         \includegraphics[width=0.9\textwidth]{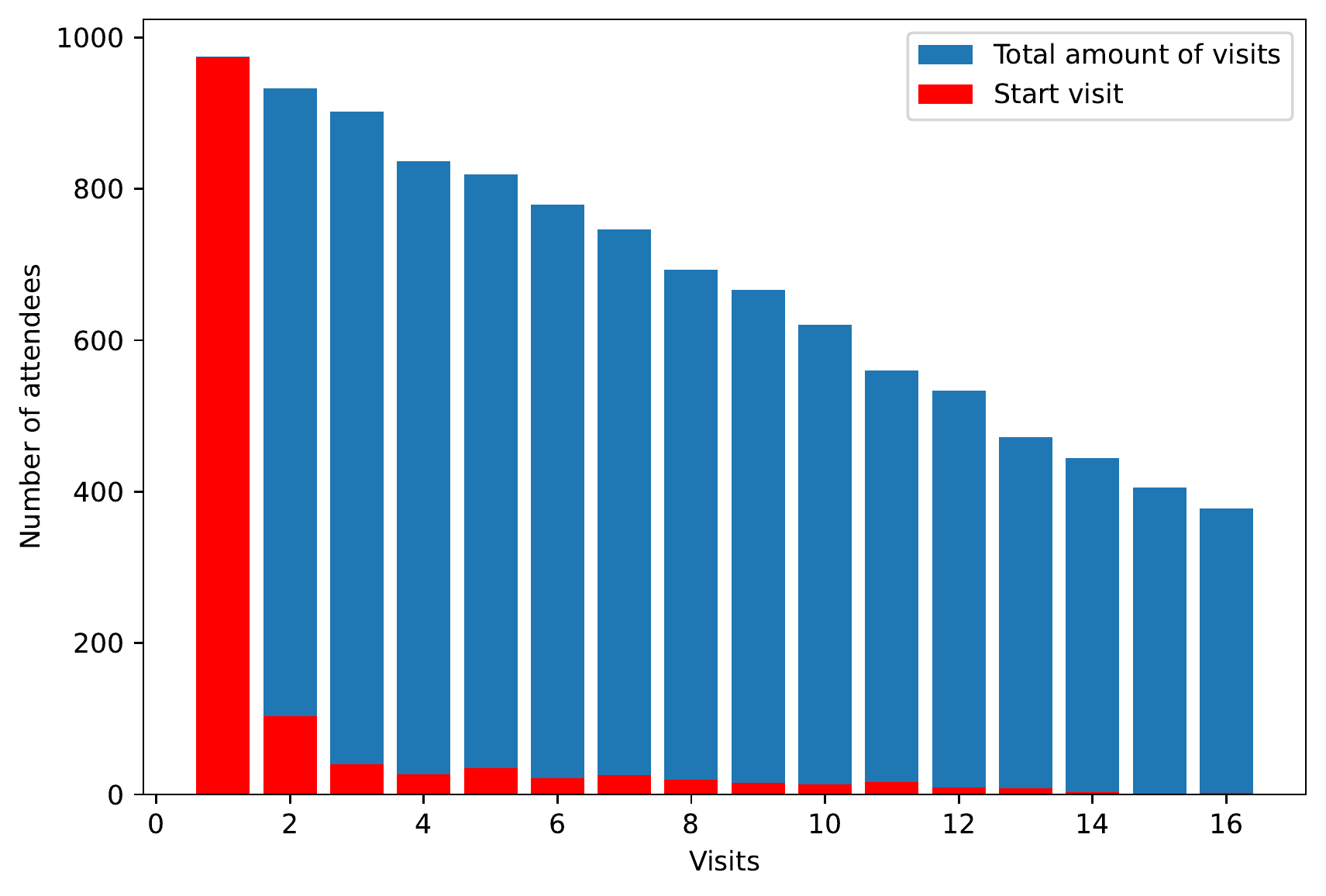}
         \caption{Attendance and entry points of specific visits}
         \label{fig:donald_missing_where}
     \end{subfigure}
    \caption{Missingness in the DONALD dataset. In (a), the amount of visits that have been attended by the participants in total can be seen. (b) depicts the number of participants that have attended a specific visit in blue and in red, the number of participants who entered the study at this specific visit is shown.}
    \label{fig:donald_missingness}
\end{figure}

Across all available records from the 1,312 participants, the dataset spans the ages three through 18, containing 36 variables. The only non-longitudinal variables are a personal number to identify each individual (pers\_ID), an identification number for each family (fam\_ID), and the sex of the participant. So a complete participant record contains 530 variables (3 static + [16 visits × 33 longitudinal variables]). Note that some participants have not been part of the study for all 16 visits, for example because they are still under 18, and that some of the yearly visits may also have been skipped for unknown reasons. Figure~\ref{fig:donald_missingness} explores the amount of missingness per participant and visit. Apart from missed visits, there is almost no missingness in the data. The only exceptions are two variables that describe the overweight status and the education level of the mother of the participants (\emph{m\_ovw} and \emph{m\_schulab}), which have small amounts of missingness (1.25\% and 0.17\%, respectively). For missing values in the original analyses (\cite{perrar_time_2019, perrar20donaldtrends}), the respective median of the total sample was used (n = 38 for maternal overweightness, n = 5 for maternal educational status).

For the application of VAMBN, the data need to be grouped into different modules. This has been initially done by experts and resulted in four modules for the DONALD data: Times (T), Nutrition (N), Anthropometric (A), and Socioeconomic (S). Over the study course, varying settings have been tested. An overview can be found in the Appendix in Table~\ref{tab:module_settings}. Moreover, for each setting, there are a few static variables -- also called covariates -- that are not grouped (i.e., the family number and the sex of the participant).     

\subsubsection{Pre-processing}
The data are stored in a tabular format with one row per visit per participant. As an identifier, the personal number is given so that each row can be assigned to a specific participant. As can be seen in Fig.~\ref{fig:donald_missingness}, the participants do not necessarily attend all 16 visits. To be able to apply the synthetic data generation methods to the DONALD data, the single rows of the dataset need to be mapped to the different visits based on the age of the participant. Thereby, we define 16 visits from zero to 15, where visit zero happens with the age of three and visit 15 with the age of 18, respectively. This results in a tabular format where each row corresponds to one single participant, including all different visits, so there are 530 columns. In this result, the degree of missingness becomes visible for every row.

\subsubsection{Post-processing}\label{subsubsec:method_post_processing}
To perform the subsequent expert analysis, we first need to convert the data back to the original format. This is done by mapping the data back to 16 rows per participant, i.e. one row per visit. Note that the synthetic data do not contain any random missingness or missed visits. 

While we are analysing the effect of sample size, we apply the following further post-processing steps to ensure that the datasets are as close as possible to the original data:
\begin{itemize}
    \item The synthetic dataset must contain the same amount of participants. 
    \item The amount of items, i.e. visits, need to correspond to the original data.
    \item The fraction of sexes in the synthetic and real cohorts should be similar. 
\end{itemize}

For the evaluation, we distinguish between raw and post-processed output to fully evaluate the algorithms (see Section~\ref{sec:evaluation_metrics} for details). Note that for both output formats the first step -- i.e. mapping the data back to 16 rows per participant -- has been applied.  

\subsection{Model}
As baseline model, we use a \textit{Variational Autoencoder Modular Bayesian Network}, short VAMBN \cite{gootjes20vambn}, which has been designed to generate fully synthetic data for longitudinal datasets containing heterogeneous features and missing values. To achieve this, a Heterogeneous-Incomplete Variational Autoencoder (HI-VAE) \cite{nazabal20hivae}, which is able to handle incomplete and heterogeneous data, is combined with a conditional Gaussian Bayesian Network (BN) \cite{heckerman_learning_1995}. To apply VAMBN, the dataset is first split into modules, which are encoded by individually trained HI-VAE modules. Thereby, different variables of the dataset are grouped together based on context, preferably together with domain experts. For each module at each time point (i.e. visit), an HI-VAE is trained, learning a low dimensional Gaussian mixture model of the input data. The resulting embeddings (consisting of discrete as well as Gaussian variables) are then used as input for a \textit{Modular Bayesian Network} (MBN), learning dependencies between these modules. At this point, auxiliary variables are introduced as missingness indicators for entire visits. For structure learning, so-called black and white lists are employed that prevent or enforce certain edges, respectively, in order to constrain the space of admissible graph structures as much as possible. From this MBN, synthetic embeddings can be drawn, and subsequently decoded by their respective HI-VAE modules to produce the final synthetic data. We refer to \cite{gootjes20vambn} for a more detailed and mathematically precise explanation.

Our extension has primarily been developed to improve VAMBN's ability to reconstruct direct mathematical dependencies between variables from different time points due to the longitudinal nature of our dataset.

\paragraph{General idea}
A simple approach to enable VAMBN to learn correlations is placing the correlated variables in the same module, because then the HI-VAE can account for their dependencies on its own. But this is limited to correlations at one specific visit, as the same variable observed at different time points is split into separate HI-VAE modules. The resulting embeddings are subsequently modelled via the BN. However, depending on the quality of the HI-VAE embeddings, this separation may result in a weakening of the temporal correlation structure after data synthesis. 

This is why in our proposed extension, all visits of each variable group are instead simultaneously encoded by one HI-VAE, which is extended by an LSTM encoder. As a result, longitudinal variables share one embedding, and thus the BN is only used to learn dependencies between the embeddings of all the different variable groups, standalone variables, and missingness indicators.

\paragraph{Architecture}
To be able to encode all $v$ visits of all $d$ variables from a module with a single HI-VAE, some changes to the pipeline have to be made. Firstly, the \textit{Evidence Lower BOund (ELBO)} -- that is used as an optimisation criterion during training of the autoencoder -- is now calculated as a sum over the contributions of all $v\cdot d$ values, to be able to account for their heterogeneous data types and varying missingness over time.

Secondly, to be able to learn dependencies over the time steps better, a Recurrent Neural Network (RNN) is inserted before the HI-VAE's recognition model (i.e., the encoder). Depending on the amount of visits in the dataset, the RNN may need to be able to represent long-term dependencies in its output. Hence, we chose a Long Short-Term Memory (LSTM) layer, due to its ability to prevent the vanishing gradients problem even when being trained on many time steps. The dimensionality of its output, which gets passed into the recognition model, is configurable. Our chosen setting can be found in the Appendix in Table~\ref{tab:hyperparameter}. 

This leads to further changes that need to be done to the generative model (i.e., the decoder): The intermediate representation vector $Y$ is now also $v$ times larger, to account for the increased number of data points encoded by the embedding $z$. Here, $Y$ is a single homogeneous intermediate representation vector, produced by a Deep Neural Network (DNN) $g(z)$ (introduced by Nazabal \textit{et al.} \cite{nazabal20hivae} in order to cope with statistical dependencies across heterogeneous data types). Note that here $g$ learns any dependencies between variables across different time points, before $Y$ is consumed by the DNNs that each parameterise one of the $v\cdot d$ distributions of attributes at specific visits. Separate parameterisations for different visits are required, since the same variable may be distributed very differently at varying time points.

The Bayesian Network still works the same as in the original VAMBN approach, although it is smaller due to the merging of all visits into one module (i.e. node of the BN).

\begin{figure}
    \centering
    \includegraphics[width=0.9\linewidth]{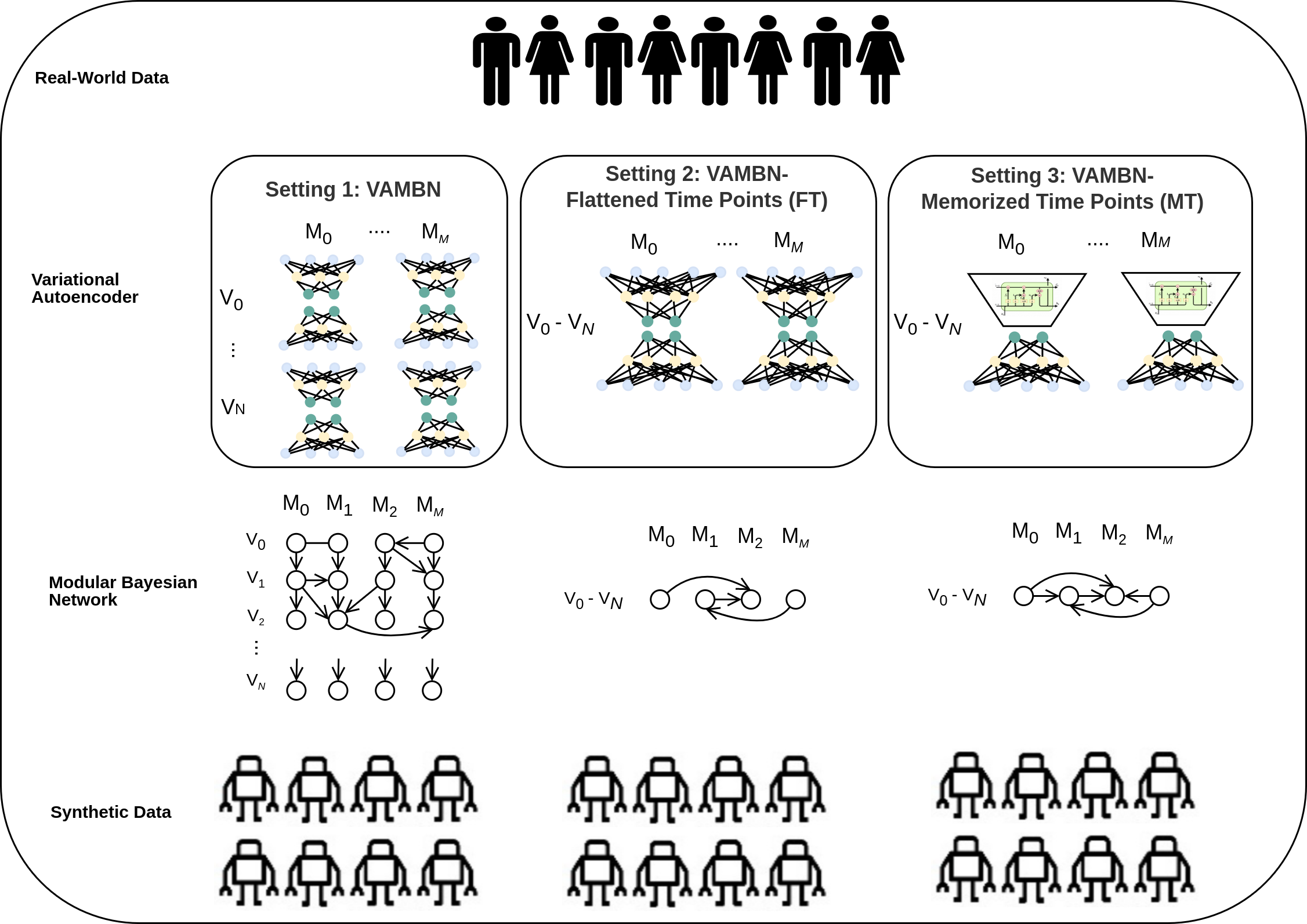}
    \caption{Overview of the developed architecture. For all settings, the real-world data are pre-processed and then embeddings are learned by an variational autoencoder (HI-VAE). Afterwards, the embeddings of the different modules are fed into a modular Bayesian Network, from which we can sample new data (that need to be decoded again by the autoencoder to be human-readable). The three shown settings differ in the neural network structure of the autoencoder (specifically, the encoder). This reduces the complexity of the Bayesian Network. In setting 2, the structure of the feedforward network is changed in such a way that all visits per module are encoded together, i.e. learned in one model. In setting 2, the default feedforward network is then changed to an LSTM layer in order to better cope with longitudinal dependencies.}
    \label{fig:developed_architecture}
\end{figure}

To analyse the effect of the two described changes in the architecture, we compare the following VAMBN variants: 
\begin{enumerate}
    \item Original VAMBN implementation
    \item VAMBN -- Memorised Time Points (VAMBN-MT): as described above, all $v$ visits of a module's $d$ variables are encoded in one HI-VAE model, its ELBO is calculated as a sum over all $v\cdot d$ contributions, and the LSTM is inserted before the encoder.
    \item VAMBN -- Flattened Time Points (VAMBN-FT): to judge the added value of the LSTM, we replace it with a standard feedforward network.
\end{enumerate}
An overview of the applied architectures can be seen in Fig.~\ref{fig:developed_architecture}.

\subsection{Evaluation Metrics}\label{sec:evaluation_metrics}
We analyse the synthetic data on different levels in order to be able to judge its quality. The methods can be divided into the following four categories -- thereof, the two latter methods are dataset-specific, since our focus lies on the utility of the generated data.

We analyse the \textbf{individual variable distributions} to ensure that they are correctly distributed across the entire synthetic population. Therefore, we provide summary statistics and density plots. Moreover, we quantify the differences between real and synthetic data distributions using the Jensen-Shannon (JS) divergence that measures the relative distance between two probability vectors \cite{e21050485}. The output ranges from 0 to 1 with 0 indicating equal distributions. Data are binned in order to get probability vectors. We use \textit{numpy's} method \textit{histogram\_bin\_edges} to determine the optimal bin size per variable by choosing the maximum of \textit{Sturges' rule} \cite{sturges_choice_1926} and the \textit{Freedman Diaconis Estimator} \cite{freedman_histogram_1981}.

Since correct distributions alone are not a sufficient indicator of convincing data however, we furthermore evaluate the reproduction of \textbf{correlations between the variables} generated by the generative model. To account for this, the Pearson correlation coefficients for all pairs of variables are calculated, including correlations across time points, and are visualised in a heatmap. For quantification, we determine the relative error $\epsilon$ of the correlation matrices as can be seen in Eq.~\ref{eq:relative_error}. Therefore, the Frobenius norm of the difference of the $real$ and $virtual$ correlation matrices is divided by the Frobenius norm of the $real$ data correlation matrix. Hence, the value can range between 0 and infinity with 0 indicating a perfect reproduction of the correlations.

\begin{equation}\label{eq:relative_error}
    \epsilon = \frac{||real-virtual||}{||real||}
\end{equation}

Even if the variable correlations are good, this still does not guarantee that the result is convincing. Variables in the original data may not only correlate, but even have direct mathematical relationships that need to be met for the synthetic data to be realistic and plausible. The existence of such \textbf{direct dependencies} is unique to the given dataset, so this work will mention them for the DONALD dataset and analyse to what degree they are met by the synthetic data.

Finally, the direct practical utility of the synthetic data can be tested by running \textbf{real-world analyses} on both the original and synthetic datasets, comparing their results. For the DONALD data, this means conducting the same time and age trend analyses in \emph{added sugar} intake following the methods by Perrar \textit{et al.} \cite{perrar_time_2019, perrar20donaldtrends}, to investigate whether we can reproduce the results. In this analysis, polynomial mixed effects regression models have been determined. We use their \emph{unadjusted models} for time and age trends. Note, that we re-built the polynomial mixed-effects models in \emph{R}, that were originally coded in \emph{SAS}, thus the values for the original data can deviate slightly from the values determined by Perrar \textit{et al.} \cite{perrar20donaldtrends}. Additionally, since our aim was not to investigate intake trends, but to generate a direct comparison between original and synthetic data, we have simplified the presentation of the trend analyses, i.e. separate presentation of age and time trends.  

\subsection{Experimental Setup}
An overview of our experimental setup can be seen in Table~\ref{tab:experimental_setup}. For the evaluation of individual variable distributions, correlations between variables, and direct dependencies, we compare our two developed methods (VAMBN-FT and VAMBN-MT) with the baseline approach (VAMBN). Because we judge their effectiveness relative to the real data, we choose the same sample size ($N=1,312$). \newline
For the subsequent real-world analyses, we choose VAMBN-MT for all experiments, but additionally investigate the influence of varying module selections -- dependent on the research question. The module selections can be found in the Appendix in Table~\ref{tab:module_settings}. For each selection, we sample 1,312 and 10,000 participants, respectively. While we want the smaller dataset to resemble the original data as much as possible, we apply the previously described post-processing (such as inserting the same percentage of missingness). In addition, we investigate the difference when using a much larger dataset without any post-processing. For each experiment and sample size, we sample 100 synthetic datasets (from the same model). To prevent influencing the time trend with outliers, we omit time points that are larger than the maximum value observed in the original data (i.e. lie in the future). For each time point, to represent all 100 trend functions,  we plot their means along with their 2.5\% and 97.5\% quantiles. \newline
For all performed experiments, we use the same black- and white lists and hyperparameter settings that can be found in the Appendix (see Table~\ref{tab:hyperparameter}).

\begin{table}[]
    \centering
    \caption{Overview of the experimental setup}
    \begin{tabular}{llrcc}
        Evaluation Method & Method & Sample Size & Modules* & Post-processed \\
        \hline
        \multirow{3}{*}{\shortstack[l]{Distributions, Correlations,\\ and Direct Dependencies}} & VAMBN & \multirow{3}{*}{1,312} & \multirow{3}{*}{(i)} & \multirow{3}{*}{No} \\
        & VAMBN-FT & &  & \\
        & VAMBN-MT & &  & \\
        \hline
        \multirow{6}{*}{Real World Analysis} & \multirow{6}{*}{VAMBN-MT} & 1,312 & \multirow{2}{*}{(i)} & Yes \\
        & & 10,000 & & No \\
        & & 1,312 & \multirow{2}{*}{(ii)} & Yes \\
        & & 10,000 & & No \\
        & & 1,312 & \multirow{2}{*}{(iii)} & Yes \\
        & & 10,000 & & No \\
        \hline
        \multicolumn{5}{l}{*See Table~\ref{tab:module_settings} for module settings.}
    \end{tabular}
    \label{tab:experimental_setup}
\end{table}

%% file: chapter/03_results.tex
\section{Results} 
We evaluated the generated synthetic data on four different levels (described in Section~\ref{sec:evaluation_metrics}). The individual variable distributions are presented in Section~\ref{subsec:results_variable_distribution}, followed by the investigation of correlations between different variables in Section~\ref{subsec:results_correlations}. Moreover, use-case specific direct dependencies are evaluated in Section~\ref{subsec:results_dependencies}. Finally, in Section~\ref{subsec:results_real_world}, the results of the real-world analyses can be found. 

\subsection{Individual Variable Distributions}\label{subsec:results_variable_distribution}
\begin{figure}
     \centering
     \begin{subfigure}[b]{0.3\textwidth}
         \centering
         \includegraphics[width=1\textwidth]{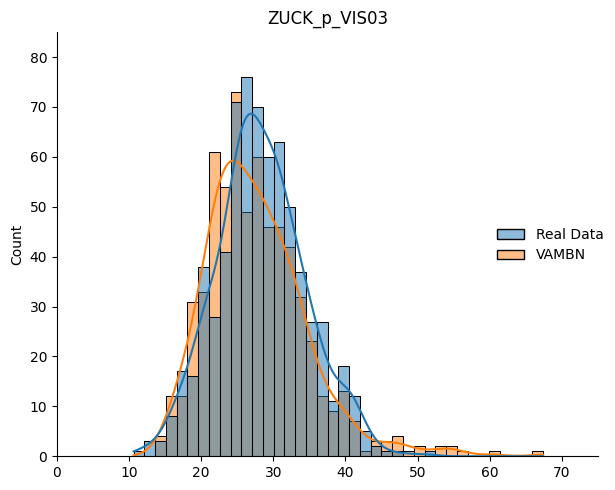}
         \caption{VAMBN}
         \label{fig:distributions_a}
     \end{subfigure}
     \begin{subfigure}[b]{0.3\textwidth}
         \centering
         \includegraphics[width=1\textwidth]{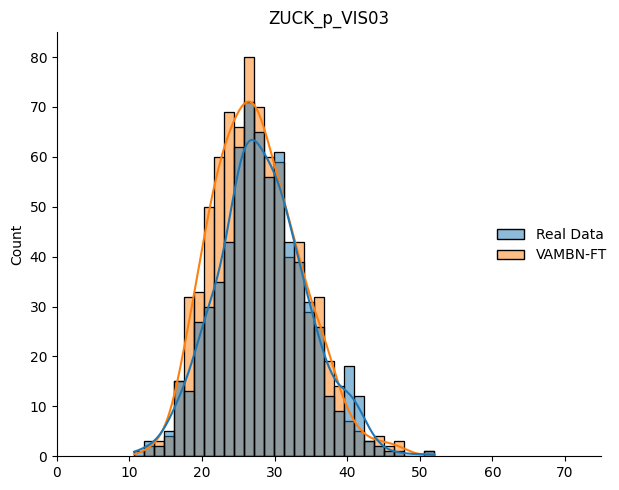}
         \caption{VAMBN-FT}
         \label{fig:distributions_b}
     \end{subfigure}
     \begin{subfigure}[b]{0.3\textwidth}
         \centering
         \includegraphics[width=1\textwidth]{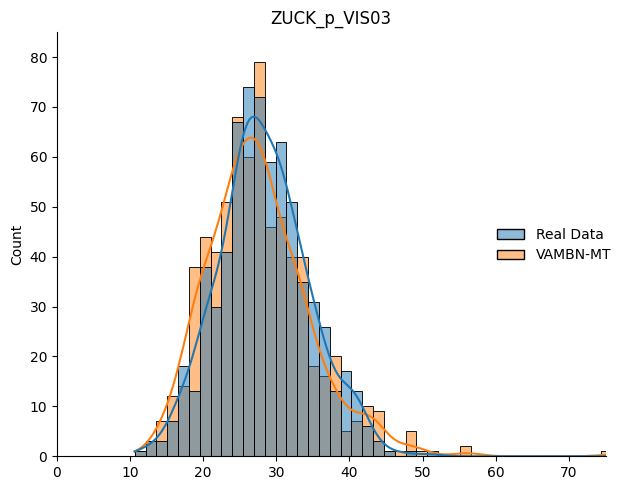}
         \caption{VAMBN-MT}
         \label{fig:distributions_b}
     \end{subfigure}
     \caption{Distribution of the total sugar intake ($ZUCK\_p$) at visit three.}
     \label{fig:distribution_zuck_p}
\end{figure}

\begin{figure}
    \captionsetup[subfigure]{justification=centering}
    \centering
     \begin{subfigure}[b]{0.3\textwidth}
         \centering
         \includegraphics[width=1\textwidth]{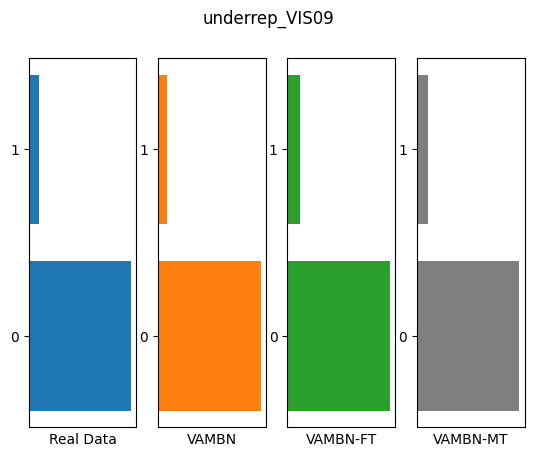}
         \caption{Distribution of underreporting \newline at visit nine}
         \label{fig:discrete_distributions_a}
     \end{subfigure}
     \begin{subfigure}[b]{0.3\textwidth}
         \centering
         \includegraphics[width=1\textwidth]{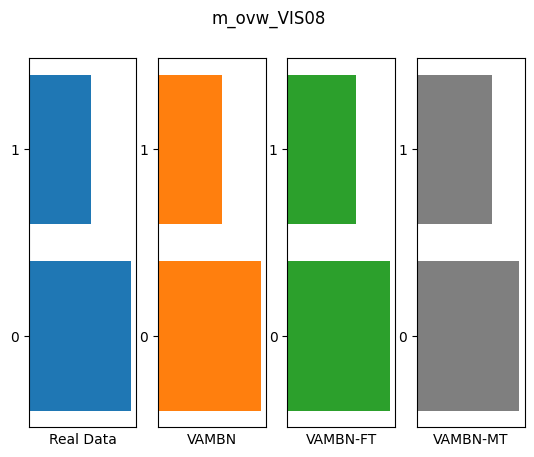}
         \caption{Overweight status of the mother \newline at visit eight}
         \label{fig:discrete_distributions_b}
     \end{subfigure}
     \begin{subfigure}[b]{0.3\textwidth}
         \centering
         \includegraphics[width=1\textwidth]{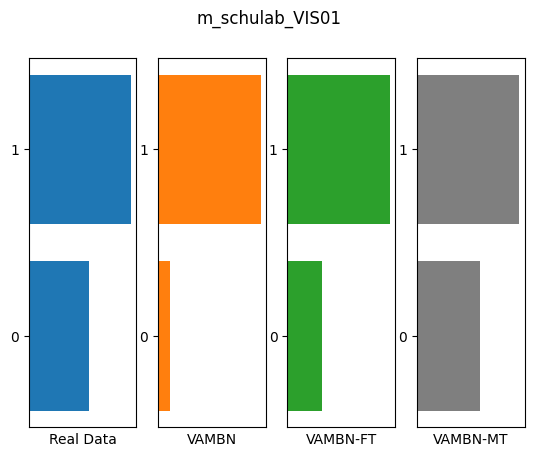}
         \caption{Distribution of the maternal education status at the first visit}
         \label{fig:discrete_distributions_c}
     \end{subfigure}
     \caption{Distribution of different discrete variables at different visits.}
     \label{fig:discrete_distribution}
\end{figure}

In order to compare the distribution between the real and synthetic data, individual distributions were plotted as a first step. Examples of these distributions can be found in Table~\ref{tab:variable_distributions}, Figure~\ref{fig:distribution_zuck_p}, and Figure~\ref{fig:discrete_distribution}, respectively. Overall, it can be seen that the summary statistics are very similar for all the different generated datasets and match the distributions of the real variables. For example, for the total sugar intake ($ZUCK\_p$), the mean of the real data amounts to 26.96, whereas the means of the synthetic data amount to 26.12, 26.1 and 26.09 for the three methods. As an example, the distribution for the third visit is visualised in Fig.~\ref{fig:distribution_zuck_p}, where slight differences between the three methods are visible and VAMBN-MT best matches the original distribution. This is also indicated by the determined JS-divergences of 0.15, 0.11 and 0.09 for the three methods, respectively. Discrete variables mostly get reconstructed correctly by VAMBN and its extensions as well. Three examples can be seen in Figure~\ref{fig:discrete_distribution}. In Fig.~\ref{fig:discrete_distributions_c}, a clear improvement can be seen for VAMBN-MT. The averaged JS-divergence over all variables (both discrete and continuous) and across all visits results in $0.15\pm0.11$ for VAMBN, in $0.12\pm 0.08$ for VAMBN-FT and $0.12\pm0.09$ for VAMBN-MT. 

\begin{table}[h]
    \centering
    \caption{Summary statistics for individual variables across all visits, covering mean, standard deviation (SD), the upper and lower quartiles (i.e. 25\% and 75\%, respectively), and the Jensen-Shannon divergence (JS-div.)}
    \begin{tabular}{llllllll}
         Variable & Dataset & Mean & SD & Median & 25\% & 75\% & JS-div. \\
         \hline
         \multirow{4}{*}{Fett\_p} & Real Data & 34.35 & 5.66 & 34.34 & 30.64 & 38.09 & -\\
         & VAMBN & 34.46 & 5.98 & 33.96 & 30.34 & 38.08 & $0.11\pm0.02$ \\
         & VAMBN-FT & 34.48 & 6.07 & 33.99 & 30.23 & 38.26 & $0.10\pm0.01$\\
         & VAMBN-MT & 34.48 & 6.08 & 34.03 & 30.26 & 38.18 & $0.11\pm0.01$ \\
         \hline
         \multirow{4}{*}{EW\_p} & Real Data & 12.99 & 2.20 & 12.85 & 11.50 & 14.27 & -\\
         & VAMBN & 13.14 & 2.18 & 12.96 & 11.63 & 14.47 & $0.09\pm0.02$ \\
         & VAMBN-FT & 13.19 & 2.23 & 13.0 & 11.64 & 14.54 & $0.09\pm0.02$\\
         & VAMBN-MT & 13.23 & 2.3 & 13.04 & 11.6 & 14.64 &  $0.09\pm0.01$\\
         \hline
         \multirow{4}{*}{ZUCK\_p} & Real Data & 26.96 & 6.72 & 26.75 & 22.37 & 31.31 & -\\
         & VAMBN & 26.12 & 5.88 & 25.69 & 22.26 & 29.28 & $0.15\pm0.02$ \\
         & VAMBN-FT & 26.1 & 6.86 & 25.51 & 21.28 & 30.27 & $0.11\pm0.02$ \\
         & VAMBN-MT & 26.09 & 7.25 & 25.63 & 20.95 & 30.55 & $0.09\pm0.02$ \\
         \hline
         \multirow{4}{*}{ZUZU\_p} & Real Data & 13.12 & 5.68 & 12.48 & 9.09 & 16.28 & -\\
         & VAMBN & 12.63 & 5.13 & 11.85 & 9.31 & 14.96 & $0.15\pm0.03$  \\
         & VAMBN-FT & 12.92 & 6.01 & 11.94 & 8.73 & 15.92 & $0.10\pm0.01$ \\
         & VAMBN-MT & 12.98 & 6.34 & 11.86 & 8.53 & 16.22 & $0.10\pm0.02$ \\
         \hline
         \multirow{4}{*}{age} & Real Data & 9.27 & 4.43 & 8.98 & 5.14 & 12.98 & -\\
         & VAMBN & 10.56 & 4.62 & 10.57 & 6.62 & 14.57 & $0.21\pm0.02$ \\
         & VAMBN-FT & 10.56 & 4.63 & 10.55 & 6.69 & 14.44 & $0.21\pm0.02$ \\
         & VAMBN-MT & 10.56 & 4.63 & 10.53 & 6.59 & 14.49 & $0.21\pm0.01$ \\
         \hline
    \end{tabular}
    \label{tab:variable_distributions}
\end{table}

\subsection{Correlations between Variables}\label{subsec:results_correlations}
\begin{figure}[h]
     \centering
     \begin{subfigure}[b]{0.34\textwidth}
         \centering
         \includegraphics[width=1\textwidth]{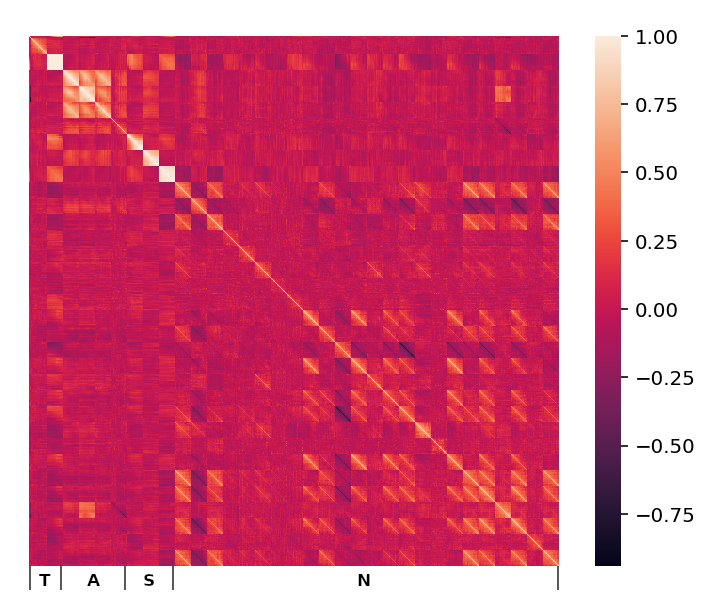}
         \caption{Original Data}
         \label{fig:correlations_a}
     \end{subfigure}
     \begin{subfigure}[b]{0.34\textwidth}
         \centering
         \includegraphics[width=1\textwidth]{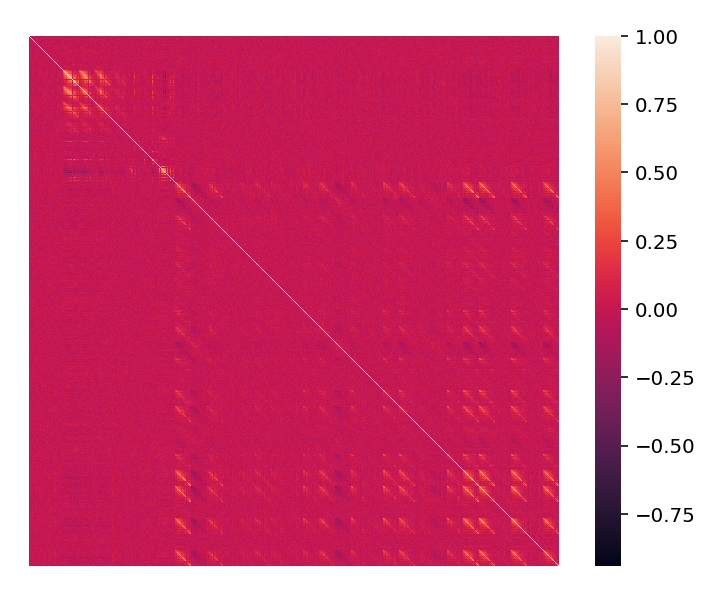}
         \caption{VAMBN $(\epsilon=0.86)$}
         \label{fig:correlations_b}
     \end{subfigure}
     \begin{subfigure}[b]{0.34\textwidth}
         \centering
         \includegraphics[width=1\textwidth]{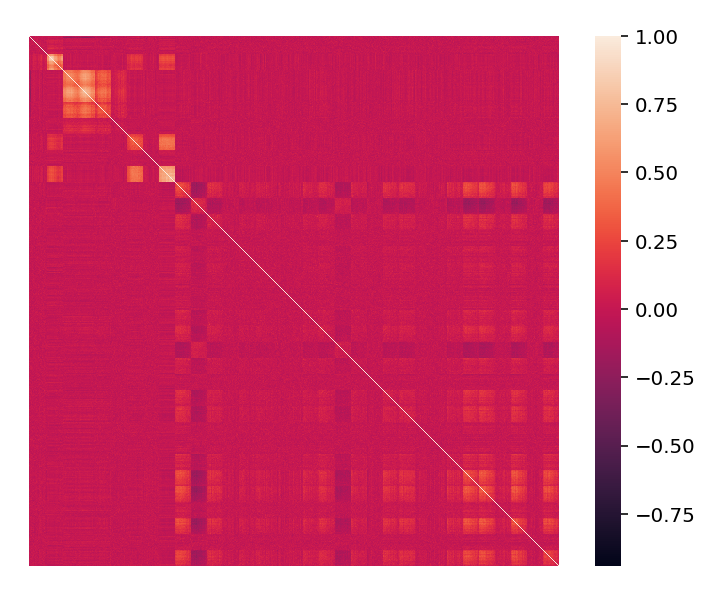}
         \caption{VAMBN-FT $(\epsilon=0.74)$}
         \label{fig:correlations_c}
     \end{subfigure}
    \begin{subfigure}[b]{0.34\textwidth}
         \centering
         \includegraphics[width=1\textwidth]{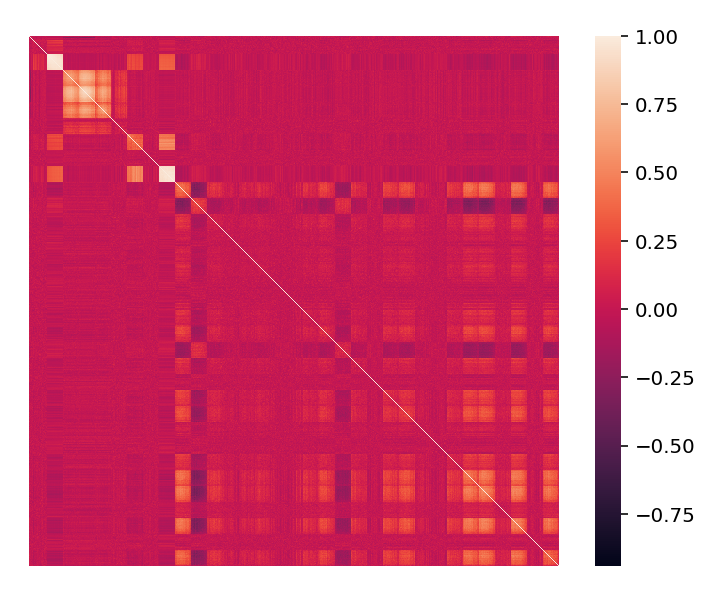}
         \caption{VAMBN-MT $(\epsilon=0.70)$}
         \label{fig:correlations_d}
     \end{subfigure}
        \caption{Heatmaps of the Pearson correlation matrices of all visits for every variable for the real data (a) and the synthetic data produced by the three different approaches (b-d). Note, that we inserted the same amount of missingness to the synthetic data to get comparable results. Their columns are sorted by variable groups (standalone > time > anthropometric > socioeconomic > nutrition, within groups alphabetical by variable name) first, indicated by the letters below (a). Within each variable, columns are sorted by visit ascending. Thus, the visible squares formed in the heatmap each correspond to the correlation of two variables over their 16 visits.}
        \label{fig:variable_correlations}
\end{figure}

We use correlations between variables as an indicator of the extent to which the properties and dependencies of the real data are represented in the synthetic data. We checked for dependencies both within the same visit and across different visits. In the heatmap representation of the Pearson correlation matrices between all variables of all visits (Figure~\ref{fig:variable_correlations}), it can be seen that absolute correlation values are generally higher in the original data (Fig.~\ref{fig:correlations_a}). With VAMBN, we obtain significantly smaller pair-wise correlations than in the real data and get a relative error of 0.86. With the VAMBN-FT version, the error is reduced to 0.74 (see Fig.~\ref{fig:correlations_c}) and the best result is achieved with VAMBN-MT with an error of 0.70 (see Fig.~\ref{fig:correlations_d}). In general, also in the original data, correlations are higher between variables of the same module than between variables of different modules, which indicates a good module selection. 

\subsection{Direct Dependencies}\label{subsec:results_dependencies}
\begin{figure}[h]
     \centering
     \begin{subfigure}[b]{0.3\textwidth}
         \centering
         \includegraphics[width=1\textwidth]{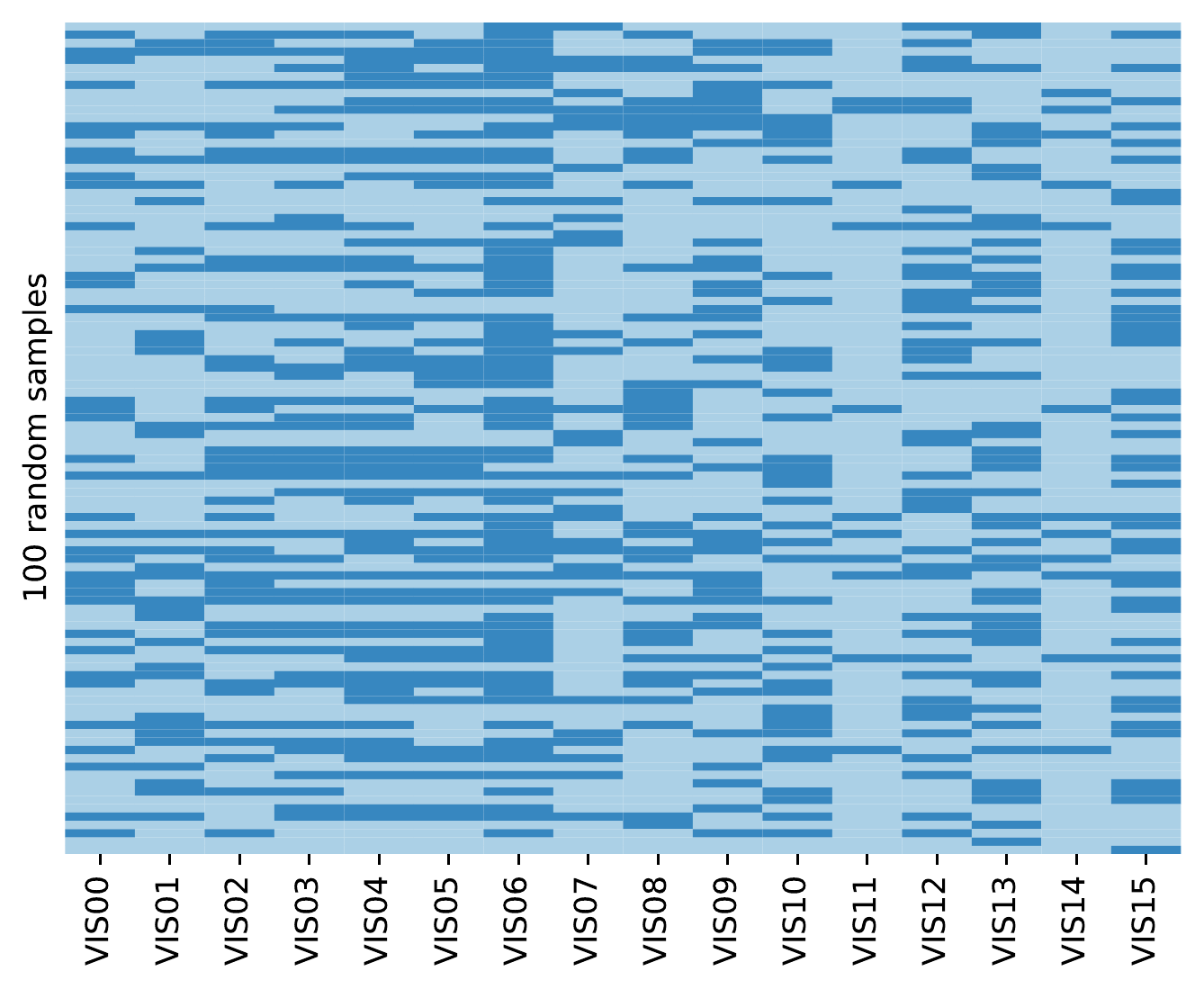}
         \caption{VAMBN}
         \label{fig:schulab_a}
     \end{subfigure}
     \begin{subfigure}[b]{0.3\textwidth}
         \centering
         \includegraphics[width=1\textwidth]{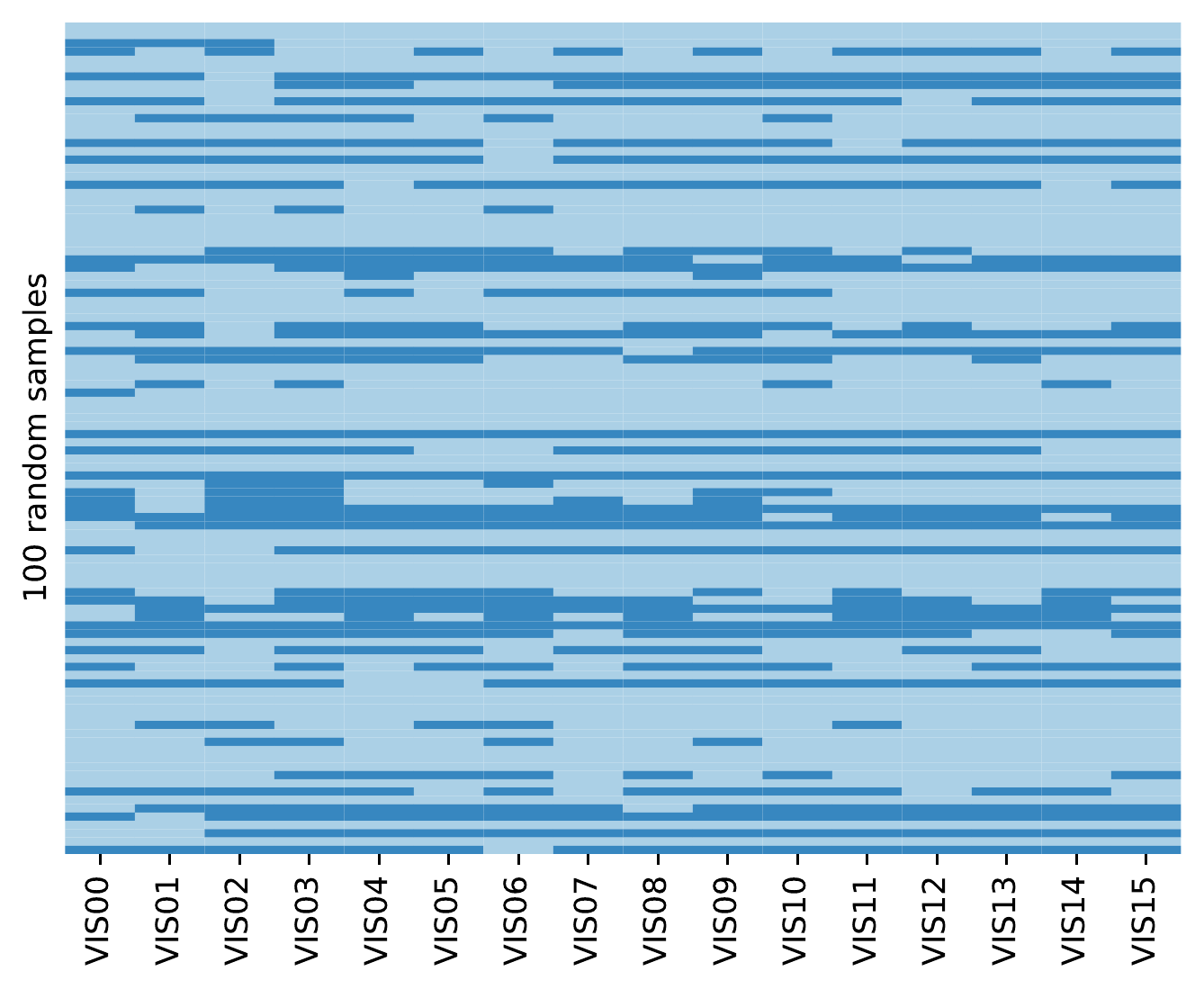}
         \caption{VAMBN-FT}
         \label{fig:schulab_b}
     \end{subfigure}
     \begin{subfigure}[b]{0.3\textwidth}
         \centering
         \includegraphics[width=1\textwidth]{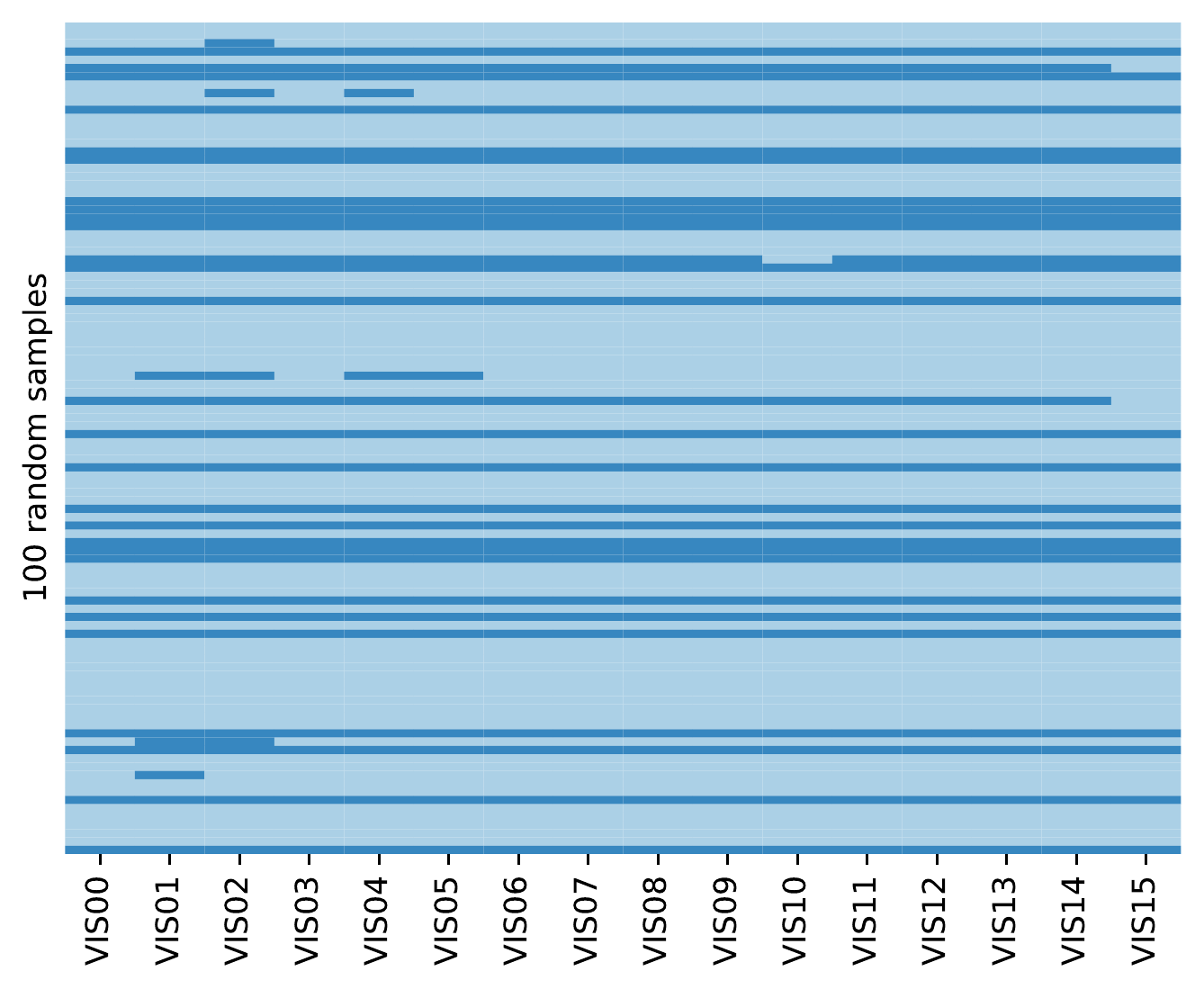}
         \caption{VAMBN-MT}
         \label{fig:schulab_c}
     \end{subfigure}
        \caption{Examples of reconstructions of the boolean variable \emph{m\_schulab} across
time. This variable indicates whether the mother of the participant has attended 12 years of school education. Whenever it was set to \textit{True} (light blue) at some point, it can logically not change back to \textit{False} (dark blue) in a later visit. The different subfigures represent the results for the three different approaches, respectively. In each case, 100 samples have been drawn randomly.}
        \label{fig:direct_dependency_schulab}
\end{figure}

The chosen DONALD dataset contains several interesting direct dependencies, representing case-specific expert knowledge, that can be used to further judge the quality of the synthetic data by investigating whether these dependencies can be correctly reproduced. Therefore, we chose two of these: Firstly, the boolean variable \emph{m\_schulab}, indicating whether the participant’s mother has had 12 years of school education. In the original dataset, this variable mostly stays the same, and might increase from 0 to 1 if a participant’s mother receives further education during the study. Logically, the indicator never changes from 1 to 0, however. 
In Figure~\ref{fig:direct_dependency_schulab}, we show the value of the graduation status over all 16 visits for 100 randomly chosen samples for each of the three experiments, respectively. It gets evident that the original VAMBN approach is not able to correctly reproduce dependencies across time, because the graduation status of the mother often changes back and forth which is not plausible (see Fig.~\ref{fig:schulab_a}). We see an improvement in the VAMBN-FT experiment (Fig.~\ref{fig:schulab_b}), which makes sense because here, the HI-VAE module receives all visits as one input, containing all variables involved in this dependency. But still, the result is not satisfactory. In contrast, only a few errors are left when using VAMBN-MT, as can be seen in Fig.~\ref{fig:schulab_c}. 

\begin{figure}
     \centering
     \begin{subfigure}[b]{0.4\textwidth}
         \centering
    \includegraphics[width=1\linewidth]{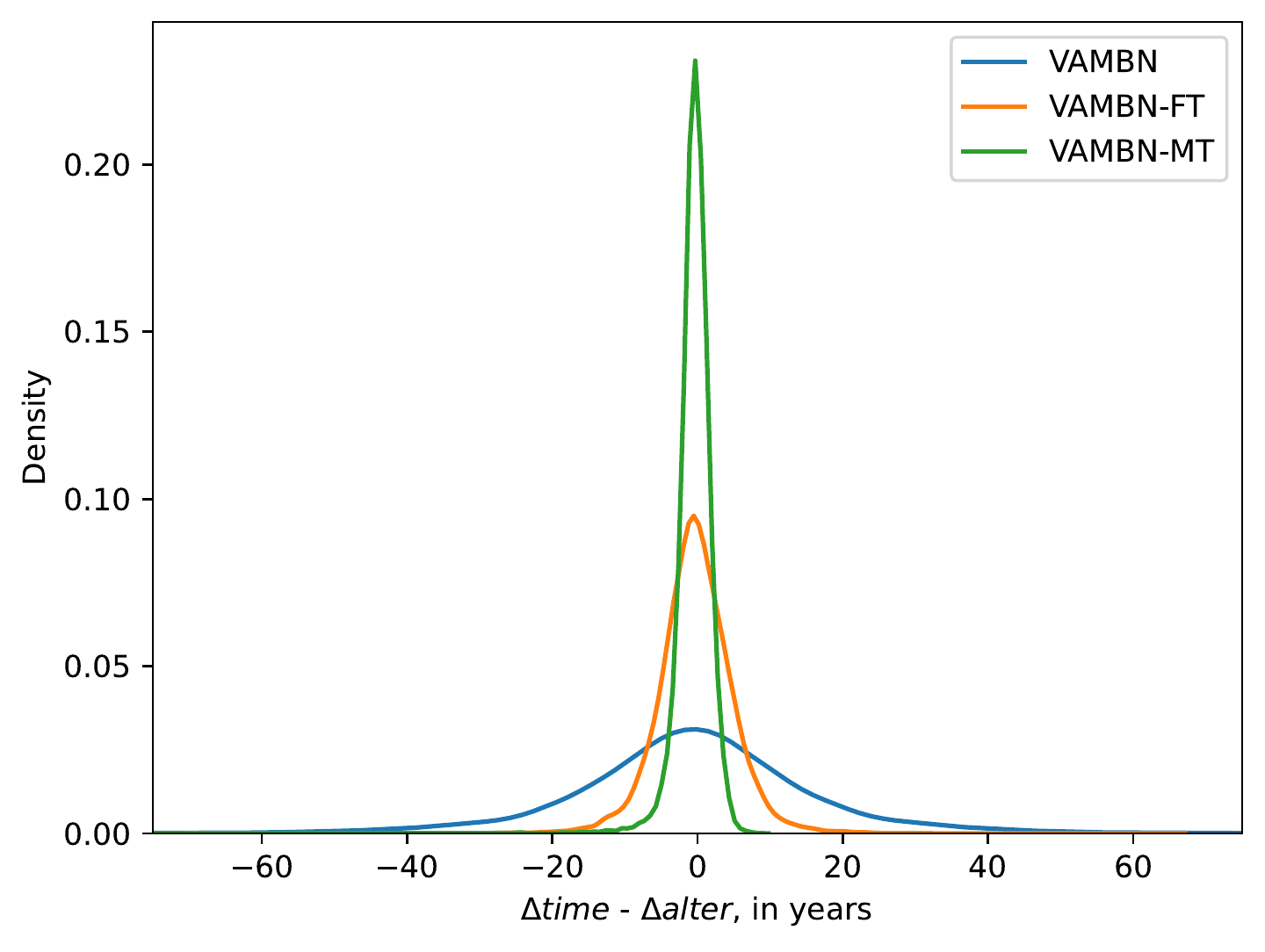}
    \caption{Error of passing time}
    \label{fig:time_error}
    \end{subfigure}
    \begin{subfigure}[b]{0.4\textwidth}
         \centering
    \includegraphics[width=1\linewidth]{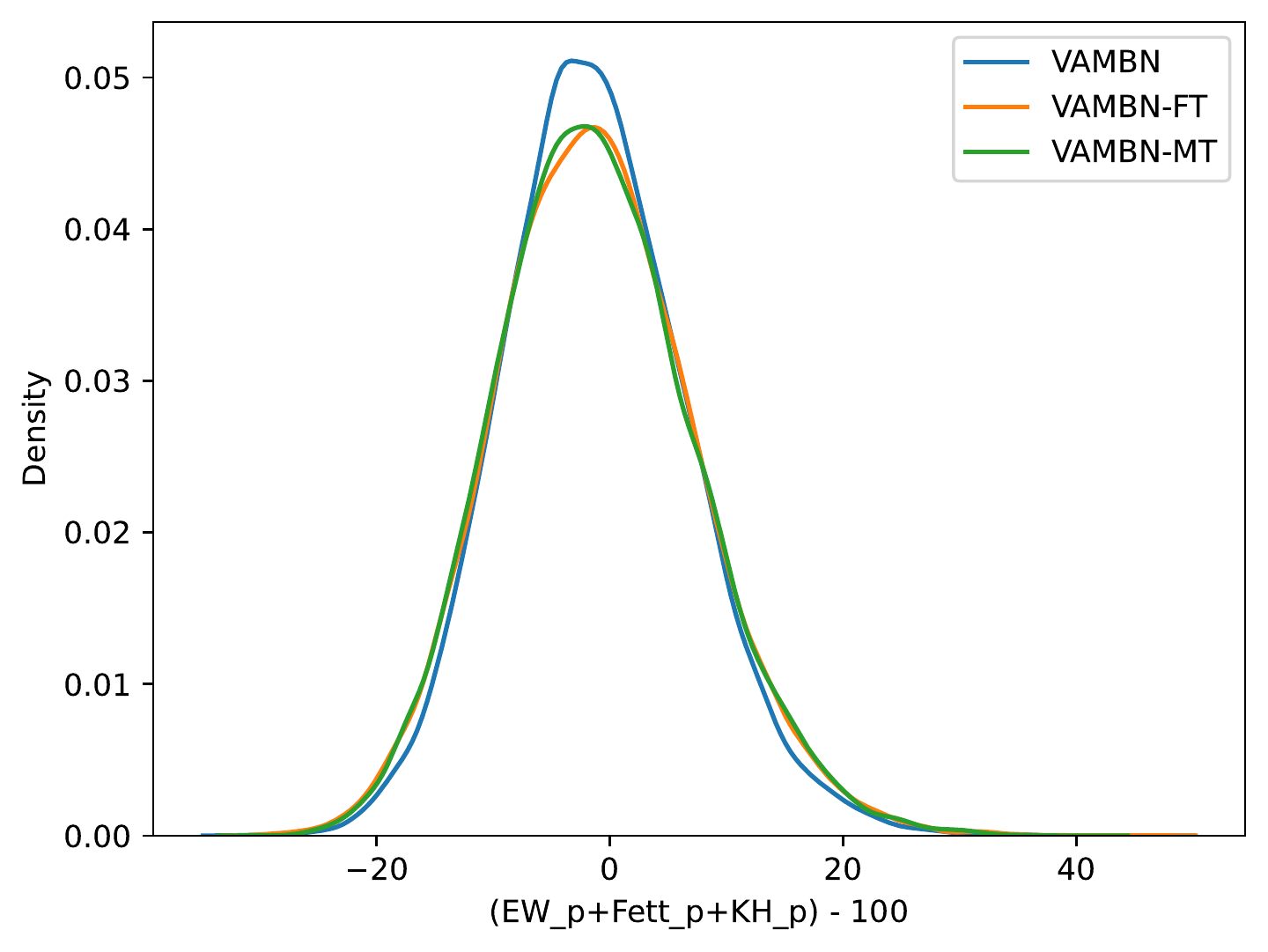}
    \caption{Summing error}
    \label{fig:100_error}
    \end{subfigure}
    \caption{Reproduction of direct mathematical dependencies. In (a), the error in passing time between the child's age and the study time is shown. It is calculated between any two successive visits, where $\Delta_{\emph{alter}}$ is the time passed for the child, and $\Delta_{\emph{time}}$ is the time passed for the study. In (b), the error in reconstructing a direct mathematical relationship within a single visit is shown, i.e. the sum of proteins (EW\_p), fats (Fett\_p) and carbohydrates (KH\_p). A perfect reconstruction of the three variables should always sum to 100.}
    \label{fig:time_and_100_error}
\end{figure}

As a second direct dependency, we investigated the relationship between the variables \emph{alter} and \emph{time}, which describe the exact age of the child and the study, respectively. They are both given as positive real-numbered values and form the variable group times. Again, for all three experiments, their summary statistics seemed plausible for all visits: The age of synthetic participants was around the expected age, with some random variance. And the age of the study was commonly anywhere between 0 and 30 years -- which is also plausible, since the study had always included children of any age over the course of its existence. When looking at an individual child over multiple visits however, the age of the study kept fluctuating, often even getting younger. Whereas in the original data, if the age has increased by $\Delta_{\emph{alter}}$, then the increase in time $\Delta_{\emph{time}}$ is exactly the same. For the virtual participants produced by the three experiments, the error $\Delta_{\emph{time}}-\Delta_{\emph{alter}}$ has been calculated for any two consecutive visits, so 15 times per participant. Figure~\ref{fig:time_error} shows the results across all visits of all participants for all three experiments. Here again, we see the worst result for the baseline approach, and the best result for VAMBN-MT. The error in passing time is clearly smaller than in the other two VAMBN versions. Hence, we can conclude that our LSTM adaption significantly improves the reconstruction of direct dependencies between variables within and across time points. 
However, these observed improvements are limited to variables across different visits with direct mathematical relationships. Similar relationships existing within a single visit are mostly unaffected by the extension, as Figure~\ref{fig:100_error} shows. 

\subsection{Real-World Analyses}\label{subsec:results_real_world}
\begin{figure}
     \centering
     \begin{subfigure}[b]{0.49\textwidth}
         \centering
    \includegraphics[width=1\linewidth]{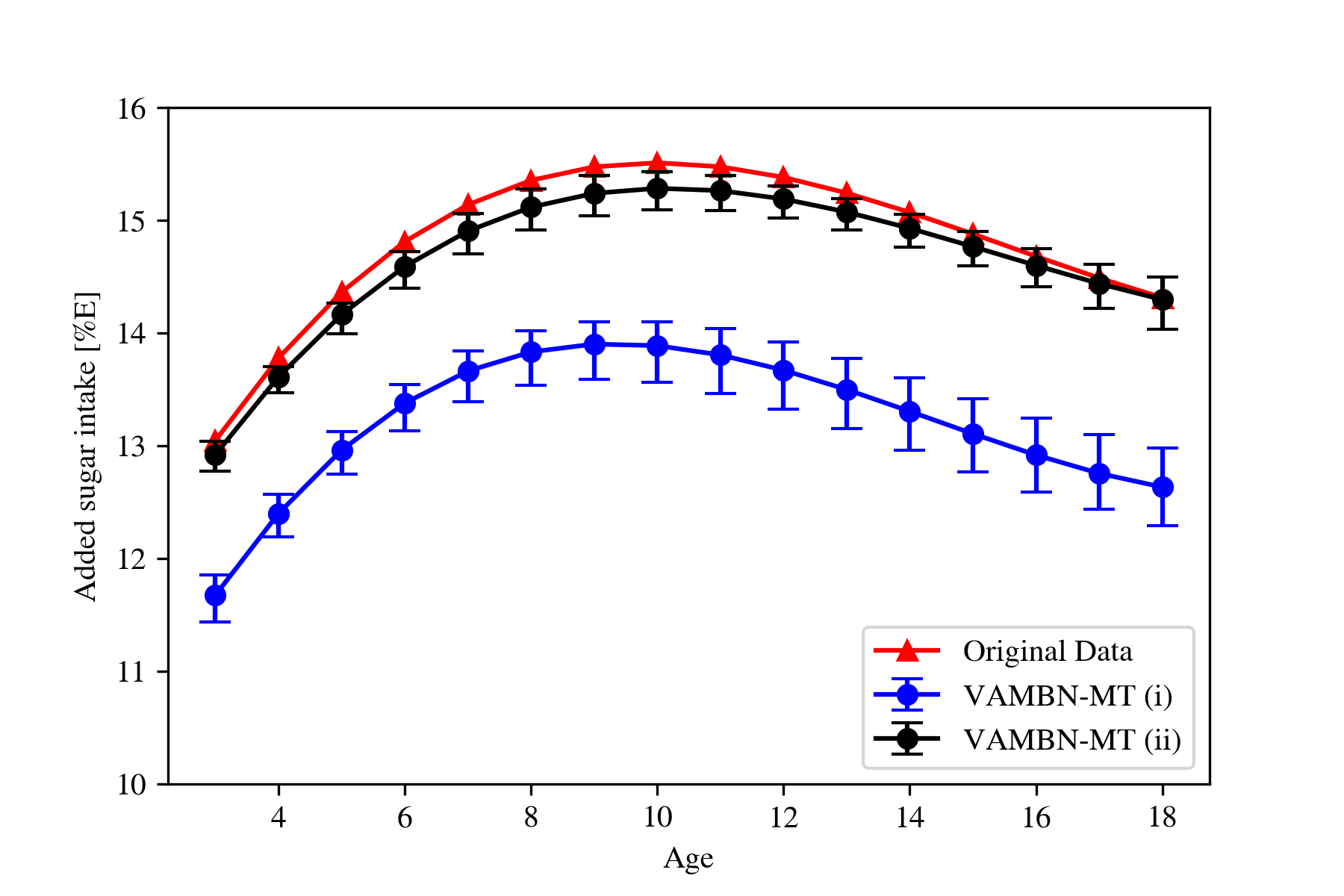}
    \caption{Age trend}
    \label{fig:age_trend_as}
    \end{subfigure}
    \begin{subfigure}[b]{0.49\textwidth}
         \centering
    \includegraphics[width=1\linewidth]{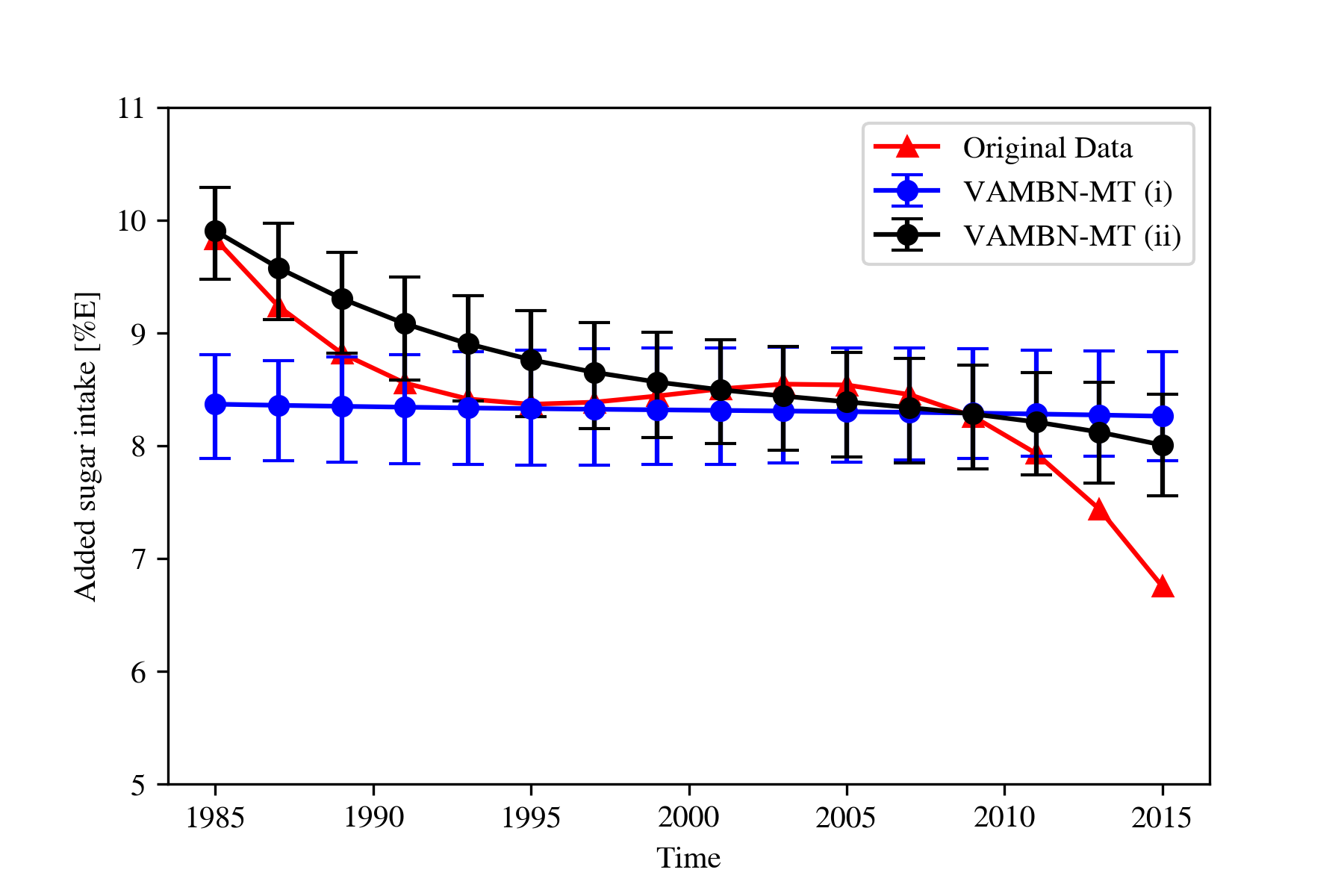}
    \caption{Time trend}
    \label{fig:time_trend_as}
    \end{subfigure}
    \caption{Comparison of age and time trends for added sugar intake predicted by polynomial mixed-effects regression models. The age and time trends of added sugar intake is shown for the real data (red) and the synthetic data (blue setting (i), and black setting (ii)) in subfigures a) and b), respectively. Whereas setting (i) corresponds to the original module selections proposed by the experts, in setting (ii) the dependent variables \emph{time}, \emph{age} and \emph{added sugar} are in one module, hence learned by the same autoencoder model. The error bar for the synthetic data indicates the confidence intervals calculated across 100 independently sampled datasets. For all synthetic datasets, the sample size is $N=10,000$.}
    \label{fig:trends_as}
\end{figure}

To finally judge the usefulness of synthetic data, we investigate and compare their performance on real-world analyses. Because VAMBN and VAMBN-FT show a lower performance in reproducing correlations between variables and direct dependencies of the dataset compared to VAMBN-MT, only VAMBN-MT's performance on real-world analyses is investigated. \newline In Figure~\ref{fig:trends_as}, the age and time trends predicted by the polynomial mixed-effects regression models for the \emph{added sugar} intake can be seen. The determined trends, including significance levels, can be found in the Appendix in Table~\ref{tab:raw_values_trend_analyses}. 

The age trend can be reproduced very well by the synthetic data and we see the same progression. Whereas the added sugar increases from the beginning up to the age of 10, it slowly decreases again afterwards. It can be seen that the variance (indicated as confidence intervals by the error bar), gets higher with increasing age. This may be due to the fact that the data basis is more incomplete for older participants. Whereas the same trend can be seen by both module selection settings (i) and (ii), selection (ii) is able to reproduce approximately same values (i.e. in terms of amplitude). In this selection, the two dependent variables \emph{age} and \emph{added sugar} are learned together in one module. Nevertheless, we achieve the same trend with module setting (i), presumably because the varying age of the participant is implicitly available to the network by the number of the visit for each module. 

The time trend can be approximated with the \emph{VAMBN-MT (ii) model} (visualised in black in  Fig.~\ref{fig:time_trend_as}). In this setting, the variables \emph{time} and \emph{added sugar} ($zuzu_p$) are learned together in one module. In contrast, the model trained with setting \emph{(i)}, where the two mentioned variables are encoded by two different autoencoders, the overall descending trend cannot be reproduced - i.e., there is no change in added sugar consumption over the years.

\textbf{Effects of sampling:} From each trained model, an infinite amount of datasets can be sampled. Whereas resampling the data lead to no visible changes in the previously shown analyses (i.e. variable distributions, correlations between variables, and direct dependencies), an effect can be seen for the real-world analyses. Here, we realised that the sample size plays an important role for the magnitude of this effect. For $N=10,000$ (without any post-processing), we see the same overall trend for all analyses, with only small variances in amplitude. However, for $N=1,312$, which corresponds to the amount of samples in the original data, and including missingness, we sometimes even experience differences in the trend. In the Appendix in Figures~\ref{fig:age_trend_as_cherry_1312} and \ref{fig:age_trend_as_cherry_10000}, examples of good and bad results can be seen for both sample sizes, respectively. 

This finding is in line with the observation that we get much higher variances in terms of confidence intervals for the smaller sample size as compared to the larger one, as can be seen in Fig.~\ref{fig:trends_as_sample_size}. Additionally, the sample size has an effect on the significance level. For the sample size of 10,000, all of the 100 sampled datasets show statistically significant results for both age and time trends -- i.e. they have p-values below 0.05 for all terms (linear, quadratic and cubic). In contrast, for the sample size of 1,312, only 70.71\% and 22.22\% of the samples produce statistically significant age and time trends, respectively. 

\textbf{Effects of real data quality:} The significance levels found in the age and time trends of the real data influence the ability to reproduce the results with synthetic data. In addition to the \emph{added sugar}, Perrar \textit{et al.} also investigated age and time trends for the \emph{total sugar} consumption. Whereas the age trends are all statistically significant (i.e. $p<0.05$), this is not the case for the time trend, where both linear and cubic terms have p-values of 0.8951 and 0.1620, respectively. This is reflected in our analyses, as we are able to reproduce the age, but not the time trend. This is summarised in Fig.~\ref{fig:trends_ts}.

%% file: chapter/04_discussion.tex
\section{Discussion}
Due to legitimate privacy concerns, access to personal health data is strictly regulated~\cite{eu16sensitivedata}, even for scientific purposes. However, the availability of data is fundamentally necessary for scientific progress. For deep learning methods to be successfully applied, it is also crucial that the available training datasets are sufficiently large \cite{deep_learning_needs_data}. To satisfy both privacy and usability demands, a classic approach is to anonymise datasets, which reduces disclosure risks and may allow sharing of the data even without informed consent. But anonymisation can severely limit the utility of medium to high-dimensional data for statistical analyses~\cite{aggarwal_k-anonymity_2005}, and there are still recent results proclaiming to be able to re-identify individuals from anonymised data~\cite{cohen-attacks, rocher_estimating_2019}. For these reasons, fully synthetic data generation methods are explored as an alternative to anonymisation. Their idea is that generating and sharing synthetic data may offer a better compromise between privacy of the individuals who appeared in the original data set, and the usability of the shared data~\cite{emam20practicalsynth}. 

A lot of research has been already performed in the area of synthetic data generation methods. Common applications are based on image, text or tabular data generation from various domains (e.g. \cite{liu_generative_2021,karras_analyzing_2020, ren_generating_2020, subramanian_adversarial_2017}). However, the heterogeneity, the incompleteness, and the longitudinality of the DONALD dataset complicate the construction of generative models that produce useful synthetic data, making the dataset particularly interesting for the development of new methods and their evaluation. Due to the challenging properties of the dataset, the majority of published methods are not suitable for the task at hand. Therefore, we chose VAMBN~\cite{gootjes20vambn} as a baseline method that is indeed designed to handle longitudinal, heterogeneous, and incomplete datasets. It does so by combining HI-VAE~\cite{nazabal20hivae}, a generative model that can encode and reconstruct heterogeneous incomplete data, with a Bayesian Network. 

In our study, we showed that VAMBN is able to reproduce summary statistics and individual variable distributions effectively. However, pair-wise correlations between different variables in the synthetic data are often not captured well (see Fig.~\ref{fig:variable_correlations}). Moreover, VAMBN fails to learn use case specific direct dependencies over time, such as the graduation status of the mother of the participants or the linear correlation between the participant's age and the time point of the study (see Figures~\ref{fig:direct_dependency_schulab} and \ref{fig:time_error}).

Therefore, we proposed an extension of VAMBN, called \emph{VAMBN-Memorised Time points} (MT), that incorporates an LSTM network, which was designed to have a long short-term memory, i.e. to better cope with time dependencies within the data. In our extension, several visits of the same variable were modelled within one module instead of splitting them into different modules that are connected via the BN. Hence, this extension firstly requires each HI-VAE module to map the data for all visits of its variable group to a single encoding, which is a process assisted by the newly introduced LSTM layer. And secondly, the extension must be able to reconstruct all data points of all visits at once from the singular encoding, which is done in a manner that ensures that all dependencies can be learned. In order to judge the improvement from including the LSTM layer, we compared both baseline and VAMBN-MT to \emph{VAMBN-Flattened Time points} (FT), which also reframes the visits so that they are learned together, but only uses the default feedforward network in the HI-VAE encoder (see Figure~\ref{fig:developed_architecture}).

In terms of pair-wise correlations between variables, VAMBN-MT clearly outperforms the two other approaches (Fig.~\ref{fig:variable_correlations}). Also the direct dependencies across visits (i.e. the education status of the mother and the linear relationship between \textit{time} and \textit{age}) are greatly improved with VAMBN-MT. An essential part of the quality analysis of synthetic data is to test whether they can be used for real-world analyses. Therefore, we use VAMBN-MT to reproduce time and age trends for added sugar intake, as done by Perrar \textit{et al.} \cite{perrar20donaldtrends}. Here, we could successfully reproduce the age trend and approximate the time trend very well (see Fig.~\ref{fig:trends_as}). Because we realised that the real-world analyses can vary between different datasets that were sampled from the same model (which is not the case for the previous analyses), we systematically determined the the effect of sampling by means of drawing new samples from each model 100 times and varying the sample size by generating datasets of size 1,312 (=size of the original data) and 10,000 (with and without post-processing, respectively). 

Thus, we draw the following conclusions: Firstly, a larger dataset leads to more stable age and time trends that only differ in amplitude but not in the progression of the trend itself (compare Fig.~\ref{fig:age_trends_as_cherrye}). Larger datasets also show less variance (in terms of confidence intervals) (Fig.~\ref{fig:trends_as_sample_size}). Moreover, whereas analyses with a small dataset lead to statistically significant results only in 22\%-70\% of cases, we get 100\% significance for the large datasets. In addition, we could show that the selection of the modules needs to be done in dependence to the research question because correlations are lost if variables are trained with separate autoencoders. Finally, we investigated the total sugar intake, that does not show statistically significant results in the real data for linear and quadratic time trends, and conclude that a non-significant trend cannot be reproduced with the synthetic data (see Figure~\ref{fig:trends_ts}).  

Altogether, we showed the importance of use case-specific evaluations incorporating expert knowledge, especially in terms of direct dependencies. With the real-world analyses, we got a demonstration of the usefulness of synthetic data and gained valuable insights into the effects of resampling and the chosen sample size, the selection of variables that need to be learned together, and the importance of the significance level of the real-world experiments.

%% file: chapter/05_conclusion.tex
\section{Conclusion and Outlook}
In this study, we generated synthetic data for a longitudinal study from the nutritional domain and performed an in-depth quality analysis, going beyond summary statistics and individual variable distributions. As we realised restrictions in the reproduction of use case specific direct dependencies across time points of current state-of-the-art methods, we developed VAMBN-MT, an LSTM-based extension of VAMBN that outperforms the original approach and is even able to reproduce real-world analyses. We highlighted the drastic increase in synthetic data quality achieved by incorporating expert domain knowledge and choosing a sufficiently large sample size. We showed that the resulting synthetic data can be a valuable source for real-world analyses. As a next step in our research, we plan to investigate the privacy risks associated with the data generated by our model to gain further insights into the risk-utility trade-off provided and to ultimately unlock the potential benefits of using synthetic data.

%% file: chapter/06_appendix.tex
\newpage
\section*{Appendix}
\setcounter{table}{0}
\renewcommand{\thetable}{A\arabic{table}}
\setcounter{figure}{0}
\renewcommand{\thefigure}{A\arabic{figure}}

\subsection*{DONALD Data}

In Table~\ref{tab:donald_metadata}, metadata for the DONALD dataset can be seen that are described in the following. Each participant gets a personal number -- a randomly generated unique identifier. In addition, each participating family gets a randomly generated family number. A family consists of the mother, the father, the biological, the adoptive, and the foster children. The sex of each participant is noted as either male or female. For every three-day weighed dietary record, the age of the participant is noted together with the date of the dietary record. Thereby, a specific time variable is created for time trend analyses, where the first included record in this evaluation was considered the baseline time, i.e., time = 0. Therefore, time ranged between 0 and 31 years.

Most variables describe the nutrient intake as a percentage of the total energy intake per day (\%E/d). This total energy intake is also given, in kilocalories per day (kcal/d).

In addition, a numerical value for the basal metabolic rate (BMR) is reported, which is the amount of calories that are burnt during rest \cite{schofield_predicting_1985}. Another variable introduced is the boolean variable of underreporting. Thereby, the plausibility of the reported values is determined based on a comparison to standard value ranges according to Sichert-Heilert \textit{et al.} \cite{sichert-hellert_underreporting_1998}. Two further measures are reported for each visit concerning the weight of the participant. The first one is the determination of obesity according to Cole \textit{et al.} (2000) and the other one is the well-known body mass index (BMI). Finally, socioeconomic factors are taken into account, consisting of the BMI of the mother of the participant, whether she is currently employed, and whether she completed her A-level (in German \textit{Fachabitur} or \textit{Abitur} - corresponding to 12 years of school education). For every three-day weighed dietary record, the number of weekdays or weekend days was recorded as well (\textit{wo\_tage}).

\begin{table}[h!] 
\centering
\caption{Overview of the variables contained within the DONALD dataset}
\begin{tabular}{lllll} 
       Variable Description & Variable name & Variable type & Unit \\
        \hline
         Personal number & pers\_ID & numerical & - \\
         Family number & fam\_ID & numerical & - \\
         Gender & sex & categorical & - \\
         \hline
         Age on first day of protocol & age & numerical & years  \\ 
         Time variable & time & numerical & years \\
         \hline
         Daily energy intake & e\_cal & numerical & kcal/d \\
         Protein intake & EW\_p & numerical & \%E/d  \\
         Fat intake & Fett\_p & numerical & \%E/d \\
         Carbohydrate intake & KH\_p & numerical & \%E/d \\
         Glucose intake & Gluc\_p & numerical & \%E/d \\
         Fructose intake & Fruc\_p & numerical & \%E/d \\
         Galactose intake & Galac\_p & numerical & \%E/d \\
         Monosaccharide intake & MSacch\_p & numerical & \%E/d \\
         Saccharose intake & Sacch\_p & numerical & \%E/d \\
         Maltose intake & MALT\_p & numerical & \%E/d \\
         Lactose intake & LACT\_p & numerical & \%E/d \\
         Disaccharide intake & DISACCH\_p & numerical & \%E/d \\
         Total sugar intake & ZUCK\_p & numerical & \%E/d \\
         Added sugar & ZUZU\_p & numerical & \%E/d \\
         Free sugar & free\_s\_p & numerical & \%E/d \\
         Free sugar from juice & fs\_saft\_p & numerical & \%E/d \\
         Free sugar from fruits and vegetables & fs\_obge\_p & numerical & \%E/d \\
         Free sugar from sugar and sweets & fs\_sp\_p & numerical & \%E/d \\
         Free sugar from bread and cake & fs\_bc\_p & numerical & \%E/d \\
         Free sugar from other sources & fs\_oth\_p & numerical & \%E/d \\
         Free sugar from dairy products & fs\_dai\_p & numerical & \%E/d \\
         Free sugar from Sugar Sweetened Beverages (SSB) & fs\_ssb\_p & numerical & \%E/d \\
         Number of weekdays & wo\_tage & categorical & - \\
         \hline
        Basal metabolic rate (BMR) & bmr & numerical & - \\
        Underreporting & underrep & categorical & - \\
        Overweight status & ovw & categorical & -  \\
        Body Mass Index (BMI) & bmi & numerical & kg/m2 \\ 
        \hline
        Overweight status of the mother & m\_ovw & categorical & - \\
        Current employment of the mother & m\_employ & categorical & - \\
        High maternal educational status of the mother & m\_schulab & categorical & - \\ 
        \hline
\end{tabular}
\label{tab:donald_metadata}
\end{table}

\subsection*{VAMBN Settings}\label{subsec:app_vambn_settings}
Table~\ref{tab:module_settings} summarises the different module selections that need to be decided before the use of VAMBN. 
\begin{table}[h!]
    \centering
    \caption{Module selections for different VAMBN-based models*.}
    \begin{tabular}{cp{2cm}p{5cm}p{2cm}p{2cm}}
         Module Setting & Time & Nutrition & Socioeconomic Factors & Anthropometric Data \\
         \hline
         (i) & age, time & e\_cal, EW\_p, Fett\_p, KH\_p, Gluc\_p, Fruc\_p, Galac\_p,
         MSacch\_p, Sacch\_p, MALT\_p, LACT\_p, DISACCH\_p, ZUCK\_p, ZUZU\_p, free\_s\_p, 
         fs\_saft\_p, fs\_obge\_p, fs\_sp\_p, fs\_bc\_p, fs\_oth\_p, fs\_dai\_p, fs\_ssb\_p, wo\_tage & 
         bmr, underrep, ovw, bmi & m\_ovw, m\_employ, m\_schulab \\ 
         \hline
         (ii) & age, time, ZUZU\_p & e\_cal, EW\_p, Fett\_p, KH\_p, Gluc\_p, Fruc\_p, Galac\_p,
         MSacch\_p, Sacch\_p, MALT\_p, LACT\_p, DISACCH\_p, ZUCK\_p, free\_s\_p, 
         fs\_saft\_p, fs\_obge\_p, fs\_sp\_p, fs\_bc\_p, fs\_oth\_p, fs\_dai\_p, fs\_ssb\_p, wo\_tage & 
         bmr, underrep, ovw, bmi & m\_ovw, m\_employ, m\_schulab \\ 
         \hline
         (ii) & age, time, ZUCK\_p & e\_cal, EW\_p, Fett\_p, KH\_p, Gluc\_p, Fruc\_p, Galac\_p,
         MSacch\_p, Sacch\_p, MALT\_p, LACT\_p, DISACCH\_p, ZUZU\_p, free\_s\_p, 
         fs\_saft\_p, fs\_obge\_p, fs\_sp\_p, fs\_bc\_p, fs\_oth\_p, fs\_dai\_p, fs\_ssb\_p, wo\_tage & 
         bmr, underrep, ovw, bmi & m\_ovw, m\_employ, m\_schulab \\
         \hline
         \multicolumn{4}{c}{*Note that fam\_ID and sex are standalone variables for each setting.}
    \end{tabular}
    \label{tab:module_settings}
\end{table}
Our black- and whitelists, which are needed for the Bayesian Network learning, contain the basic constraints that an edge can never go back in time, and that the standalone variables cannot depend on any other variable. \newline
Moreover, the used hyperparameters for training can be found in Table~\ref{tab:hyperparameter}. 

\begin{table}[h!]
    \centering
    \caption{Hyperparameters used for HI-VAE training}
    \begin{tabular}{lll}
         Hyperparameter & Module & Value \\
         \hline
         \multirow{4}{*}{Learning rate}  & Times & \multirow{4}{*}{0.01} \\
         & Nutrition & \\
         & Anthropometric & \\
         & Socioeconomic & \\
         \hline
         \multirow{4}{*}{Batch Size} & Times & \multirow{4}{*}{127} \\
         & Nutrition & \\
         & Anthropometric & \\
         & Socioeconomic & \\
         \hline
         \multirow{4}{*}{Y-Dimensionality} & Times & \multirow{4}{*}{1} \\
         & Nutrition & \\
         & Anthropometric & \\
         & Socioeconomic & \\
         \hline
         \multirow{4}{*}{S-Dimensionality} & Times & \multirow{2}{*}{1} \\
         & Nutrition & \\
         & Anthropometric &  \multirow{2}{*}{2}\\
         & Socioeconomic & \\
         \hline
         \multirow{4}{*}{LSTM-Dimensionality} & Times & \multirow{4}{*}{20} \\
         & Nutrition & \\
         & Anthropometric & \\
         & Socioeconomic & \\
         \hline
    \end{tabular}
    \label{tab:hyperparameter}
\end{table}

\subsection*{Real-world Analyses}
\begin{landscape}
\begin{small}
\begin{longtable}{lllllll}
    \caption{Results of polynomial mixed-effect models for age and time trends of added sugar intake for the different generated datasets} \\
    Method & $Age$ $(p)$ & $Age^2$ $(p)$ & $Age^3$ $(p)$ & $Time$ $(p)$ & $Time^2$ $(p)$ & $Time^3$ $(p)$ \\
    \hline
    \endfirsthead
     Method & $Age$ $(p)$ & $Age^2$ $(p)$ & $Age^3$ $(p)$ & $Time$ $(p)$ & $Time^2$ $(p)$ & $Time^3$ $(p)$ \\
     \hline
     \endhead
     Original Data* & 1.350827 (<0.0001) & -0.099685 (<0.0001) & 0.002137 (0.0002) & -0.0.348131 (<0.0001) & 0.026152 (<0.0001) & -0.000599 (<0.0001) \\
     \hline
    \multirow{49}{*}{VAMBN-MT (i) 10000}& 1.430963 (<0.0001) & -0.115705 (<0.0001) & 0.002723 (<0.0001) & 0.025684 (0.223012) & -0.001424 (0.333201) & 2.8e-05 (0.370199) \\
    & 1.408156 (<0.0001) & -0.111171 (<0.0001) & 0.002584 (<0.0001) & -0.003557 (0.864749) & 4e-06 (0.998019) & -2e-06 (0.945519) \\
    & 1.393149 (<0.0001) & -0.112023 (<0.0001) & 0.002664 (<0.0001) & 0.024316 (0.244942) & -0.002527 (0.083921) & 6e-05 (0.049591) \\
    & 1.342718 (<0.0001) & -0.105812 (<0.0001) & 0.002428 (<0.0001) & -0.003396 (0.87088) & 0.000347 (0.812486) & -1.3e-05 (0.683313) \\
    & 1.38336 (<0.0001) & -0.108266 (<0.0001) & 0.002462 (<0.0001) & -0.012054 (0.561269) & 0.001444 (0.320452) & -4e-05 (0.18968) \\
    & 1.476569 (<0.0001) & -0.118884 (<0.0001) & 0.00281 (<0.0001) & -0.005234 (0.798662) & 7.9e-05 (0.956205) & -4e-06 (0.893813) \\
    & 1.417386 (<0.0001) & -0.113722 (<0.0001) & 0.002656 (<0.0001) & 0.004955 (0.812989) & -0.000329 (0.822024) & 1e-05 (0.746472) \\
    & 1.440692 (<0.0001) & -0.11426 (<0.0001) & 0.002631 (<0.0001) & 0.014392 (0.479681) & -0.001513 (0.291507) & 3.5e-05 (0.242204) \\
    & 1.52464 (<0.0001) & -0.123319 (<0.0001) & 0.002921 (<0.0001) & -0.005849 (0.779298) & 0.000798 (0.582831) & -2.1e-05 (0.491004) \\
    & 1.478183 (<0.0001) & -0.121326 (<0.0001) & 0.002939 (<0.0001) & -0.01 (0.623962) & 0.00043 (0.764083) & -7e-06 (0.828089) \\
    & 1.416625 (<0.0001) & -0.113158 (<0.0001) & 0.002644 (<0.0001) & -0.009194 (0.65675) & 0.0007 (0.628279) & -1.3e-05 (0.679093) \\
    & 1.468699 (<0.0001) & -0.116248 (<0.0001) & 0.002683 (<0.0001) & -0.021168 (0.303979) & 0.000888 (0.540096) & -1.3e-05 (0.666859) \\
    & 1.362627 (<0.0001) & -0.110525 (<0.0001) & 0.002614 (<0.0001) & 0.024577 (0.228469) & -0.000575 (0.687768) & -2e-06 (0.958185) \\
    & 1.429001 (<0.0001) & -0.113265 (<0.0001) & 0.002622 (<0.0001) & -0.021336 (0.306943) & 0.001751 (0.229002) & -4.4e-05 (0.152824) \\
    & 1.421166 (<0.0001) & -0.115259 (<0.0001) & 0.002743 (<0.0001) & 0.021191 (0.303541) & -0.002611 (0.070101) & 6.7e-05 (0.027548) \\
    & 1.447328 (<0.0001) & -0.114806 (<0.0001) & 0.002693 (<0.0001) & 0.007654 (0.71308) & -0.001309 (0.368884) & 2.5e-05 (0.411436) \\
    & 1.365656 (<0.0001) & -0.106593 (<0.0001) & 0.002449 (<0.0001) & -0.035603 (0.08799) & 0.002134 (0.144344) & -5.1e-05 (0.098494) \\
    & 1.399197 (<0.0001) & -0.109818 (<0.0001) & 0.0025 (<0.0001) & -0.01189 (0.565467) & 0.000805 (0.578397) & -2.2e-05 (0.465431) \\
    & 1.387137 (<0.0001) & -0.108345 (<0.0001) & 0.002413 (<0.0001) & -0.001731 (0.933152) & 0.000121 (0.933097) & 1e-06 (0.972541) \\
    & 1.358621 (<0.0001) & -0.104747 (<0.0001) & 0.002335 (<0.0001) & -0.03694 (0.072523) & 0.001696 (0.238648) & -2.5e-05 (0.403834) \\
    & 1.428979 (<0.0001) & -0.113885 (<0.0001) & 0.002644 (<0.0001) & -0.011523 (0.582107) & 0.000795 (0.586429) & -2.1e-05 (0.487305) \\
    & 1.4598 (<0.0001) & -0.119574 (<0.0001) & 0.002879 (<0.0001) & -0.005296 (0.798037) & 0.000549 (0.705309) & -9e-06 (0.77996) \\
    & 1.386755 (<0.0001) & -0.109129 (<0.0001) & 0.002485 (<0.0001) & -0.002027 (0.921946) & 0.000281 (0.846035) & -1.2e-05 (0.697522) \\
    & 1.386846 (<0.0001) & -0.1092 (<0.0001) & 0.002499 (<0.0001) & -0.021206 (0.30399) & 0.001618 (0.261885) & -4.2e-05 (0.170433) \\
    & 1.335938 (<0.0001) & -0.104791 (<0.0001) & 0.00239 (<0.0001) & 0.00188 (0.928299) & -0.00016 (0.913141) & 1e-06 (0.964105) \\
    & 1.357753 (<0.0001) & -0.105788 (<0.0001) & 0.002385 (<0.0001) & 0.039295 (0.058432) & -0.003142 (0.030688) & 6.4e-05 (0.036516) \\
    & 1.461898 (<0.0001) & -0.114953 (<0.0001) & 0.002655 (<0.0001) & -0.02713 (0.187143) & 0.001143 (0.427771) & -2.4e-05 (0.435993) \\
    & 1.407884 (<0.0001) & -0.110518 (<0.0001) & 0.002541 (<0.0001) & 0.020444 (0.322373) & -0.002323 (0.108) & 5.1e-05 (0.094491) \\
    & 1.317254 (<0.0001) & -0.102333 (<0.0001) & 0.002291 (<0.0001) & 0.027303 (0.188922) & -0.00276 (0.057939) & 6.4e-05 (0.034668) \\
    & 1.345957 (<0.0001) & -0.105435 (<0.0001) & 0.002385 (<0.0001) & 0.008335 (0.685253) & -0.000206 (0.886198) & -1e-06 (0.981259) \\
    & 1.293585 (<0.0001) & -0.102001 (<0.0001) & 0.002362 (<0.0001) & -0.031494 (0.122066) & 0.002192 (0.126255) & -5.6e-05 (0.06444) \\
    & 1.475885 (<0.0001) & -0.118306 (<0.0001) & 0.002808 (<0.0001) & -0.023357 (0.255611) & 0.001337 (0.353625) & -3e-05 (0.318661) \\
    & 1.428994 (<0.0001) & -0.111113 (<0.0001) & 0.002519 (<0.0001) & -0.025981 (0.210783) & 0.001134 (0.433863) & -2.2e-05 (0.473848) \\
    & 1.476203 (<0.0001) & -0.120321 (<0.0001) & 0.002882 (<0.0001) & 0.002931 (0.887087) & -0.000183 (0.898464) & 3e-06 (0.922449) \\
    & 1.474939 (<0.0001) & -0.121196 (<0.0001) & 0.002974 (<0.0001) & -0.020884 (0.31299) & 0.001194 (0.409684) & -2.7e-05 (0.372887) \\
    & 1.418087 (<0.0001) & -0.111836 (<0.0001) & 0.002565 (<0.0001) & -0.021492 (0.298061) & 0.001728 (0.233093) & -4.2e-05 (0.164551) \\
    & 1.444998 (<0.0001) & -0.114757 (<0.0001) & 0.002667 (<0.0001) & -0.023323 (0.267852) & 0.000866 (0.55408) & -1.1e-05 (0.716039) \\
    & 1.478548 (<0.0001) & -0.11785 (<0.0001) & 0.002737 (<0.0001) & -0.060341 (0.003639) & 0.0045 (0.00194) & -9.1e-05 (0.003019) \\
    & 1.382281 (<0.0001) & -0.108553 (<0.0001) & 0.002463 (<0.0001) & -0.043568 (0.036158) & 0.003166 (0.029088) & -6.3e-05 (0.037765) \\
    & 1.494474 (<0.0001) & -0.120808 (<0.0001) & 0.002887 (<0.0001) & -0.016763 (0.423466) & 0.00079 (0.588707) & -1.3e-05 (0.68002) \\
    & 1.470761 (<0.0001) & -0.118528 (<0.0001) & 0.002815 (<0.0001) & -0.018833 (0.371642) & 0.001263 (0.390297) & -2.8e-05 (0.365204) \\
    & 1.458587 (<0.0001) & -0.117286 (<0.0001) & 0.002782 (<0.0001) & -0.013594 (0.509744) & 0.000526 (0.716379) & -1.4e-05 (0.640502) \\
    & 1.441864 (<0.0001) & -0.117848 (<0.0001) & 0.002818 (<0.0001) & -0.011221 (0.586412) & 0.000199 (0.889935) & 1.2e-05 (0.6951) \\
    & 1.413548 (<0.0001) & -0.113437 (<0.0001) & 0.002682 (<0.0001) & 0.013876 (0.49674) & -0.001498 (0.296447) & 2.6e-05 (0.394443) \\
    & 1.346716 (<0.0001) & -0.106522 (<0.0001) & 0.002469 (<0.0001) & 0.022395 (0.280621) & -0.001447 (0.320242) & 2.1e-05 (0.499586) \\
    & 1.345751 (<0.0001) & -0.103189 (<0.0001) & 0.002278 (<0.0001) & -0.002915 (0.887453) & -0.000768 (0.592851) & 2.6e-05 (0.391435) \\
    & 1.368118 (<0.0001) & -0.107463 (<0.0001) & 0.002459 (<0.0001) & -0.005478 (0.790552) & 3.3e-05 (0.981523) & 1e-06 (0.976322) \\
    & 1.480253 (<0.0001) & -0.118944 (<0.0001) & 0.002794 (<0.0001) & -0.010069 (0.623167) & 0.000196 (0.891376) & 6e-06 (0.830876) \\
    & 1.326192 (<0.0001) & -0.102482 (<0.0001) & 0.002287 (<0.0001) & -0.004953 (0.808166) & -0.000245 (0.864172) & 9e-06 (0.775369) \\
    & 1.351115 (<0.0001) & -0.10755 (<0.0001) & 0.002507 (<0.0001) & 0.013192 (0.517679) & -0.001217 (0.393365) & 2.5e-05 (0.399862) \\
    & 1.483649 (<0.0001) & -0.121139 (<0.0001) & 0.002912 (<0.0001) & 0.012019 (0.560765) & -0.000902 (0.533781) & 2.1e-05 (0.486449) \\
    & 1.446879 (<0.0001) & -0.118278 (<0.0001) & 0.002837 (<0.0001) & -0.012258 (0.556064) & 0.001572 (0.280706) & -3.6e-05 (0.237239) \\
    & 1.527557 (<0.0001) & -0.123504 (<0.0001) & 0.002962 (<0.0001) & -0.042967 (0.038767) & 0.00265 (0.067918) & -5.2e-05 (0.084573) \\
    & 1.43545 (<0.0001) & -0.112875 (<0.0001) & 0.002586 (<0.0001) & -0.034076 (0.097255) & 0.002055 (0.153274) & -3.9e-05 (0.197503) \\
    & 1.303301 (<0.0001) & -0.100605 (<0.0001) & 0.002249 (<0.0001) & 0.00865 (0.673972) & -0.001382 (0.335993) & 3e-05 (0.319191) \\
    & 1.405868 (<0.0001) & -0.108531 (<0.0001) & 0.0024 (<0.0001) & -0.031472 (0.132698) & 0.00254 (0.081823) & -5.3e-05 (0.081686) \\
    & 1.399375 (<0.0001) & -0.110676 (<0.0001) & 0.002564 (<0.0001) & -0.01149 (0.581785) & 0.000313 (0.830542) & 1e-06 (0.973605) \\
    & 1.281933 (<0.0001) & -0.103361 (<0.0001) & 0.002438 (<0.0001) & 0.020085 (0.326038) & -0.000782 (0.585134) & 8e-06 (0.799232) \\
    & 1.349634 (<0.0001) & -0.10585 (<0.0001) & 0.002403 (<0.0001) & -0.004155 (0.838867) & 0.000322 (0.821998) & -4e-06 (0.904015) \\
    & 1.364358 (<0.0001) & -0.106072 (<0.0001) & 0.002388 (<0.0001) & 0.008029 (0.699222) & -0.000873 (0.546937) & 1e-05 (0.739627) \\
    & 1.443567 (<0.0001) & -0.115641 (<0.0001) & 0.002706 (<0.0001) & -0.012288 (0.554651) & 0.001244 (0.39238) & -3.2e-05 (0.289797) \\
    & 1.454254 (<0.0001) & -0.11577 (<0.0001) & 0.002693 (<0.0001) & -0.001591 (0.939025) & -0.000473 (0.744761) & 1e-05 (0.739129) \\
    & 1.444097 (<0.0001) & -0.114522 (<0.0001) & 0.002655 (<0.0001) & 0.031836 (0.117793) & -0.00284 (0.048347) & 6e-05 (0.050229) \\
    & 1.442224 (<0.0001) & -0.114217 (<0.0001) & 0.00264 (<0.0001) & 0.000895 (0.966115) & -0.000541 (0.712784) & 1e-05 (0.747102) \\
    & 1.490825 (<0.0001) & -0.119925 (<0.0001) & 0.002843 (<0.0001) & 0.031324 (0.126492) & -0.002277 (0.113367) & 4.3e-05 (0.159138) \\
    & 1.324253 (<0.0001) & -0.100786 (<0.0001) & 0.002203 (<0.0001) & -0.020895 (0.314023) & 0.001074 (0.460326) & -2.6e-05 (0.403222) \\
    & 1.426621 (<0.0001) & -0.111815 (<0.0001) & 0.002569 (<0.0001) & -0.011499 (0.584375) & 0.000481 (0.743915) & -1.6e-05 (0.593483) \\
    & 1.414004 (<0.0001) & -0.11348 (<0.0001) & 0.002679 (<0.0001) & 0.003697 (0.85685) & -0.000937 (0.512887) & 2.8e-05 (0.356533) \\
    & 1.43872 (<0.0001) & -0.116193 (<0.0001) & 0.002741 (<0.0001) & 0.024766 (0.224126) & -0.001858 (0.194665) & 3.9e-05 (0.19801) \\
    & 1.298988 (<0.0001) & -0.098369 (<0.0001) & 0.002123 (<0.0001) & 0.008495 (0.677172) & -0.000362 (0.800355) & 4e-06 (0.895529) \\
    & 1.387075 (<0.0001) & -0.110072 (<0.0001) & 0.002567 (<0.0001) & -0.016088 (0.435804) & 0.001107 (0.442614) & -2.5e-05 (0.400936) \\
    & 1.509474 (<0.0001) & -0.12257 (<0.0001) & 0.002931 (<0.0001) & -0.006449 (0.752771) & 0.001017 (0.478511) & -2.7e-05 (0.369954) \\
    & 1.470529 (<0.0001) & -0.118254 (<0.0001) & 0.002779 (<0.0001) & -0.036158 (0.077618) & 0.001852 (0.197812) & -2.5e-05 (0.405801) \\
    & 1.573805 (<0.0001) & -0.128209 (<0.0001) & 0.003086 (<0.0001) & -0.010065 (0.622273) & 0.000445 (0.755962) & -1e-06 (0.970441) \\
    & 1.406657 (<0.0001) & -0.112822 (<0.0001) & 0.002651 (<0.0001) & 0.01043 (0.616199) & -0.000788 (0.5885) & 1.6e-05 (0.599419) \\
    & 1.370864 (<0.0001) & -0.107385 (<0.0001) & 0.002408 (<0.0001) & 0.005866 (0.776976) & 0.00036 (0.802976) & -1.3e-05 (0.657442) \\
    & 1.523548 (<0.0001) & -0.126305 (<0.0001) & 0.00308 (<0.0001) & 0.001538 (0.940116) & 0.000577 (0.687226) & -1.7e-05 (0.565294) \\
    & 1.378437 (<0.0001) & -0.108065 (<0.0001) & 0.002455 (<0.0001) & -0.034645 (0.097555) & 0.002258 (0.121906) & -4.5e-05 (0.138035) \\
    & 1.544045 (<0.0001) & -0.125286 (<0.0001) & 0.002976 (<0.0001) & -0.004757 (0.818638) & 0.000608 (0.675014) & -1.5e-05 (0.633871) \\
    & 1.503052 (<0.0001) & -0.120792 (<0.0001) & 0.002857 (<0.0001) & -0.042589 (0.040859) & 0.002571 (0.076724) & -4.8e-05 (0.11748) \\
    & 1.390467 (<0.0001) & -0.110119 (<0.0001) & 0.002533 (<0.0001) & -0.002266 (0.912069) & -0.000231 (0.872262) & 9e-06 (0.768442) \\
    & 1.414249 (<0.0001) & -0.112108 (<0.0001) & 0.00262 (<0.0001) & -0.003567 (0.861562) & -5.4e-05 (0.970161) & -1e-05 (0.750992) \\
    & 1.469364 (<0.0001) & -0.117569 (<0.0001) & 0.002751 (<0.0001) & -0.008028 (0.694807) & 0.000363 (0.800273) & -1e-05 (0.737872) \\
    & 1.299703 (<0.0001) & -0.098335 (<0.0001) & 0.002127 (<0.0001) & -0.009952 (0.628129) & 0.000137 (0.924038) & -5e-06 (0.873261) \\
    & 1.410711 (<0.0001) & -0.113418 (<0.0001) & 0.002681 (<0.0001) & 0.007177 (0.72768) & -0.000763 (0.597231) & 1.6e-05 (0.609111) \\
    & 1.338628 (<0.0001) & -0.104918 (<0.0001) & 0.002371 (<0.0001) & -0.01837 (0.37159) & 0.00088 (0.541495) & -1.1e-05 (0.708367) \\
    & 1.388275 (<0.0001) & -0.110323 (<0.0001) & 0.00257 (<0.0001) & -0.004447 (0.828335) & -0.000603 (0.675696) & 2.2e-05 (0.471391) \\
    & 1.392872 (<0.0001) & -0.11121 (<0.0001) & 0.002604 (<0.0001) & 0.022404 (0.280169) & -0.002197 (0.129836) & 4.9e-05 (0.111219) \\
    & 1.42927 (<0.0001) & -0.111038 (<0.0001) & 0.002509 (<0.0001) & -0.016409 (0.428503) & 0.000273 (0.849898) & -2e-06 (0.944615) \\
    & 1.386567 (<0.0001) & -0.109781 (<0.0001) & 0.002538 (<0.0001) & -0.000783 (0.969331) & -0.000255 (0.858578) & 1e-06 (0.968916) \\
    & 1.35645 (<0.0001) & -0.109237 (<0.0001) & 0.00257 (<0.0001) & 0.001233 (0.952657) & 0.000754 (0.60269) & -2.3e-05 (0.449388) \\
    & 1.425244 (<0.0001) & -0.111522 (<0.0001) & 0.002513 (<0.0001) & -0.022883 (0.270143) & 0.000763 (0.598114) & -8e-06 (0.798353) \\
    & 1.459754 (<0.0001) & -0.117174 (<0.0001) & 0.002752 (<0.0001) & -0.009631 (0.645646) & 0.001346 (0.358105) & -3.6e-05 (0.244771) \\
    & 1.445619 (<0.0001) & -0.116337 (<0.0001) & 0.002743 (<0.0001) & 0.016156 (0.441381) & -0.001039 (0.477864) & 2e-05 (0.52013) \\
    & 1.380316 (<0.0001) & -0.108038 (<0.0001) & 0.002443 (<0.0001) & -0.01311 (0.527251) & 0.000851 (0.557448) & -1.8e-05 (0.559665) \\
    & 1.44928 (<0.0001) & -0.116439 (<0.0001) & 0.00274 (<0.0001) & -0.005309 (0.800387) & 0.001204 (0.41241) & -3.8e-05 (0.219339) \\
    & 1.374484 (<0.0001) & -0.108541 (<0.0001) & 0.002489 (<0.0001) & -0.015462 (0.448542) & 0.001058 (0.458205) & -2.5e-05 (0.408646) \\
    & 1.48552 (<0.0001) & -0.119793 (<0.0001) & 0.002831 (<0.0001) & 0.012325 (0.548489) & -0.0006 (0.675096) & 2e-06 (0.943282) \\
    & 1.436821 (<0.0001) & -0.114275 (<0.0001) & 0.002644 (<0.0001) & 0.004479 (0.82618) & -0.000806 (0.573507) & 1.6e-05 (0.589287) \\
    & 1.407821 (<0.0001) & -0.109305 (<0.0001) & 0.00247 (<0.0001) & -0.009089 (0.661229) & -0.00079 (0.585617) & 2.7e-05 (0.371808) \\
     \hline
     \multirow{34}{*}{VAMBN-MT (ii) 10000} & 1.197875 (<0.0001) & -0.087233 (<0.0001) & 0.001873 (<0.0001) & -0.196814 (<0.0001) & 0.009335 (<0.0001) & -0.000166 (<0.0001) \\
    & 1.140093 (<0.0001) & -0.080435 (<0.0001) & 0.001626 (<0.0001) & -0.159657 (<0.0001) & 0.006642 (<0.0001) & -0.00011 (0.000538) \\
    & 1.142855 (<0.0001) & -0.080905 (<0.0001) & 0.001656 (<0.0001) & -0.161226 (<0.0001) & 0.005816 (<0.0001) & -8.3e-05 (0.008958) \\
    & 1.310469 (<0.0001) & -0.096164 (<0.0001) & 0.002084 (<0.0001) & -0.16336 (<0.0001) & 0.006711 (<0.0001) & -0.000111 (0.000556) \\
    & 1.228712 (<0.0001) & -0.087582 (<0.0001) & 0.00182 (<0.0001) & -0.17659 (<0.0001) & 0.007702 (<0.0001) & -0.000132 (<0.0001) \\
    & 1.299052 (<0.0001) & -0.097819 (<0.0001) & 0.002179 (<0.0001) & -0.158307 (<0.0001) & 0.006751 (<0.0001) & -0.000118 (0.000212) \\
    & 1.223881 (<0.0001) & -0.088469 (<0.0001) & 0.001873 (<0.0001) & -0.174889 (<0.0001) & 0.007812 (<0.0001) & -0.000137 (<0.0001) \\
    & 1.314663 (<0.0001) & -0.097038 (<0.0001) & 0.002112 (<0.0001) & -0.2144 (<0.0001) & 0.010245 (<0.0001) & -0.00018 (<0.0001) \\
    & 1.342093 (<0.0001) & -0.098732 (<0.0001) & 0.002132 (<0.0001) & -0.179686 (<0.0001) & 0.007656 (<0.0001) & -0.000127 (<0.0001) \\
    & 1.246543 (<0.0001) & -0.08966 (<0.0001) & 0.001877 (<0.0001) & -0.191126 (<0.0001) & 0.007977 (<0.0001) & -0.000125 (<0.0001) \\
    & 1.186373 (<0.0001) & -0.086792 (<0.0001) & 0.001883 (<0.0001) & -0.160864 (<0.0001) & 0.006351 (<0.0001) & -9.6e-05 (0.002788) \\
    & 1.222457 (<0.0001) & -0.08976 (<0.0001) & 0.001926 (<0.0001) & -0.155478 (<0.0001) & 0.006767 (<0.0001) & -0.00012 (0.000189) \\
    & 1.283121 (<0.0001) & -0.095167 (<0.0001) & 0.002105 (<0.0001) & -0.194908 (<0.0001) & 0.008616 (<0.0001) & -0.000147 (<0.0001) \\
    & 1.294416 (<0.0001) & -0.096725 (<0.0001) & 0.002139 (<0.0001) & -0.159968 (<0.0001) & 0.006477 (<0.0001) & -0.000103 (0.001378) \\
    & 1.264627 (<0.0001) & -0.092908 (<0.0001) & 0.002007 (<0.0001) & -0.205591 (<0.0001) & 0.009556 (<0.0001) & -0.000164 (<0.0001) \\
    & 1.211946 (<0.0001) & -0.087647 (<0.0001) & 0.00186 (<0.0001) & -0.166185 (<0.0001) & 0.006521 (<0.0001) & -9.5e-05 (0.002908) \\
    & 1.137246 (<0.0001) & -0.078739 (<0.0001) & 0.001557 (<0.0001) & -0.185759 (<0.0001) & 0.008095 (<0.0001) & -0.000138 (<0.0001) \\
    & 1.336356 (<0.0001) & -0.098988 (<0.0001) & 0.002161 (<0.0001) & -0.191472 (<0.0001) & 0.008663 (<0.0001) & -0.000148 (<0.0001) \\
    & 1.315905 (<0.0001) & -0.095873 (<0.0001) & 0.002035 (<0.0001) & -0.19204 (<0.0001) & 0.008293 (<0.0001) & -0.000138 (<0.0001) \\
    & 1.255946 (<0.0001) & -0.092612 (<0.0001) & 0.00203 (<0.0001) & -0.174545 (<0.0001) & 0.007035 (<0.0001) & -0.000113 (0.000464) \\
    & 1.418519 (<0.0001) & -0.10828 (<0.0001) & 0.002466 (<0.0001) & -0.173419 (<0.0001) & 0.006532 (<0.0001) & -9.6e-05 (0.00202) \\
    & 1.37659 (<0.0001) & -0.106386 (<0.0001) & 0.002482 (<0.0001) & -0.189723 (<0.0001) & 0.008941 (<0.0001) & -0.000161 (<0.0001) \\
    & 1.265672 (<0.0001) & -0.093292 (<0.0001) & 0.002011 (<0.0001) & -0.156296 (<0.0001) & 0.006241 (<0.0001) & -9.6e-05 (0.002459) \\
    & 1.276042 (<0.0001) & -0.094103 (<0.0001) & 0.002041 (<0.0001) & -0.14568 (<0.0001) & 0.005661 (0.000161) & -9.1e-05 (0.004708) \\
    & 1.139085 (<0.0001) & -0.080644 (<0.0001) & 0.001658 (<0.0001) & -0.171934 (<0.0001) & 0.007919 (<0.0001) & -0.000146 (<0.0001) \\
    & 1.231197 (<0.0001) & -0.088917 (<0.0001) & 0.001884 (<0.0001) & -0.187278 (<0.0001) & 0.008428 (<0.0001) & -0.000146 (<0.0001) \\
    & 1.266312 (<0.0001) & -0.094724 (<0.0001) & 0.0021 (<0.0001) & -0.19022 (<0.0001) & 0.008875 (<0.0001) & -0.000158 (<0.0001) \\
    & 1.268357 (<0.0001) & -0.093894 (<0.0001) & 0.002051 (<0.0001) & -0.211777 (<0.0001) & 0.010461 (<0.0001) & -0.000189 (<0.0001) \\
    & 1.29376 (<0.0001) & -0.095616 (<0.0001) & 0.002093 (<0.0001) & -0.194139 (<0.0001) & 0.008349 (<0.0001) & -0.000132 (<0.0001) \\
    & 1.254384 (<0.0001) & -0.091245 (<0.0001) & 0.001955 (<0.0001) & -0.1931 (<0.0001) & 0.008603 (<0.0001) & -0.000143 (<0.0001) \\
    & 1.420929 (<0.0001) & -0.106224 (<0.0001) & 0.002346 (<0.0001) & -0.199873 (<0.0001) & 0.00934 (<0.0001) & -0.00016 (<0.0001) \\
    & 1.224548 (<0.0001) & -0.087484 (<0.0001) & 0.001827 (<0.0001) & -0.183936 (<0.0001) & 0.008275 (<0.0001) & -0.000143 (<0.0001) \\
    & 1.289497 (<0.0001) & -0.0967 (<0.0001) & 0.002133 (<0.0001) & -0.160829 (<0.0001) & 0.00649 (<0.0001) & -0.000101 (0.001291) \\
    & 1.271421 (<0.0001) & -0.093913 (<0.0001) & 0.002055 (<0.0001) & -0.18703 (<0.0001) & 0.008627 (<0.0001) & -0.000155 (<0.0001) \\
    & 1.311954 (<0.0001) & -0.098954 (<0.0001) & 0.002231 (<0.0001) & -0.165116 (<0.0001) & 0.006875 (<0.0001) & -0.000114 (0.000403) \\
    & 1.317976 (<0.0001) & -0.098538 (<0.0001) & 0.002171 (<0.0001) & -0.175963 (<0.0001) & 0.00782 (<0.0001) & -0.00013 (<0.0001) \\
    & 1.271006 (<0.0001) & -0.093908 (<0.0001) & 0.002041 (<0.0001) & -0.147005 (<0.0001) & 0.005296 (0.000326) & -8e-05 (0.010291) \\
    & 1.288979 (<0.0001) & -0.091541 (<0.0001) & 0.001884 (<0.0001) & -0.184931 (<0.0001) & 0.008111 (<0.0001) & -0.000135 (<0.0001) \\
    & 1.415764 (<0.0001) & -0.106503 (<0.0001) & 0.002372 (<0.0001) & -0.198219 (<0.0001) & 0.008766 (<0.0001) & -0.000146 (<0.0001) \\
    & 1.213079 (<0.0001) & -0.087217 (<0.0001) & 0.001839 (<0.0001) & -0.189393 (<0.0001) & 0.007878 (<0.0001) & -0.000127 (<0.0001) \\
    & 1.142052 (<0.0001) & -0.080149 (<0.0001) & 0.001605 (<0.0001) & -0.15201 (<0.0001) & 0.006751 (<0.0001) & -0.000123 (<0.0001) \\
    & 1.402261 (<0.0001) & -0.107284 (<0.0001) & 0.00246 (<0.0001) & -0.17991 (<0.0001) & 0.007561 (<0.0001) & -0.000121 (0.000167) \\
    & 1.290111 (<0.0001) & -0.094875 (<0.0001) & 0.00205 (<0.0001) & -0.200172 (<0.0001) & 0.009331 (<0.0001) & -0.000164 (<0.0001) \\
    & 1.357088 (<0.0001) & -0.103223 (<0.0001) & 0.002343 (<0.0001) & -0.206863 (<0.0001) & 0.009772 (<0.0001) & -0.000169 (<0.0001) \\
    & 1.106836 (<0.0001) & -0.078241 (<0.0001) & 0.001594 (<0.0001) & -0.192259 (<0.0001) & 0.008746 (<0.0001) & -0.00015 (<0.0001) \\
    & 1.303812 (<0.0001) & -0.098496 (<0.0001) & 0.002214 (<0.0001) & -0.183922 (<0.0001) & 0.008632 (<0.0001) & -0.000156 (<0.0001) \\
    & 1.201231 (<0.0001) & -0.085696 (<0.0001) & 0.001765 (<0.0001) & -0.193588 (<0.0001) & 0.008456 (<0.0001) & -0.000138 (<0.0001) \\
    & 1.190787 (<0.0001) & -0.087772 (<0.0001) & 0.0019 (<0.0001) & -0.183968 (<0.0001) & 0.007817 (<0.0001) & -0.000122 (0.000119) \\
    & 1.278096 (<0.0001) & -0.094334 (<0.0001) & 0.002053 (<0.0001) & -0.195062 (<0.0001) & 0.00917 (<0.0001) & -0.000163 (<0.0001) \\
    & 1.272487 (<0.0001) & -0.094134 (<0.0001) & 0.002052 (<0.0001) & -0.177653 (<0.0001) & 0.007301 (<0.0001) & -0.000114 (0.000268) \\
    & 1.265323 (<0.0001) & -0.094337 (<0.0001) & 0.002081 (<0.0001) & -0.173451 (<0.0001) & 0.007676 (<0.0001) & -0.000132 (<0.0001) \\
    & 1.280096 (<0.0001) & -0.09472 (<0.0001) & 0.002076 (<0.0001) & -0.20432 (<0.0001) & 0.009496 (<0.0001) & -0.000164 (<0.0001) \\
    & 1.324767 (<0.0001) & -0.096675 (<0.0001) & 0.002079 (<0.0001) & -0.180672 (<0.0001) & 0.007714 (<0.0001) & -0.00013 (<0.0001) \\
    & 1.271809 (<0.0001) & -0.094074 (<0.0001) & 0.002042 (<0.0001) & -0.158978 (<0.0001) & 0.006643 (<0.0001) & -0.000112 (0.000493) \\
    & 1.415406 (<0.0001) & -0.108458 (<0.0001) & 0.002491 (<0.0001) & -0.17482 (<0.0001) & 0.00677 (<0.0001) & -9.8e-05 (0.002064) \\
    & 1.213286 (<0.0001) & -0.087185 (<0.0001) & 0.001808 (<0.0001) & -0.15224 (<0.0001) & 0.00592 (<0.0001) & -9.5e-05 (0.002324) \\
    & 1.315186 (<0.0001) & -0.096216 (<0.0001) & 0.002072 (<0.0001) & -0.238751 (<0.0001) & 0.011864 (<0.0001) & -0.000209 (<0.0001) \\
    & 1.192898 (<0.0001) & -0.083992 (<0.0001) & 0.00169 (<0.0001) & -0.185915 (<0.0001) & 0.008143 (<0.0001) & -0.000135 (<0.0001) \\
    & 1.200757 (<0.0001) & -0.086666 (<0.0001) & 0.001829 (<0.0001) & -0.169705 (<0.0001) & 0.007251 (<0.0001) & -0.000126 (<0.0001) \\
    & 1.298997 (<0.0001) & -0.097433 (<0.0001) & 0.002151 (<0.0001) & -0.189486 (<0.0001) & 0.008285 (<0.0001) & -0.000137 (<0.0001) \\
    & 1.353886 (<0.0001) & -0.102239 (<0.0001) & 0.002293 (<0.0001) & -0.186484 (<0.0001) & 0.008792 (<0.0001) & -0.000156 (<0.0001) \\
    & 1.351991 (<0.0001) & -0.103287 (<0.0001) & 0.002339 (<0.0001) & -0.168583 (<0.0001) & 0.007063 (<0.0001) & -0.000116 (0.000265) \\
    & 1.236619 (<0.0001) & -0.090211 (<0.0001) & 0.001924 (<0.0001) & -0.194118 (<0.0001) & 0.008972 (<0.0001) & -0.000154 (<0.0001) \\
    & 1.201102 (<0.0001) & -0.086973 (<0.0001) & 0.001841 (<0.0001) & -0.193079 (<0.0001) & 0.008736 (<0.0001) & -0.000148 (<0.0001) \\
    & 1.319192 (<0.0001) & -0.098446 (<0.0001) & 0.002186 (<0.0001) & -0.169736 (<0.0001) & 0.007168 (<0.0001) & -0.000121 (0.000162) \\
    & 1.188878 (<0.0001) & -0.084317 (<0.0001) & 0.001699 (<0.0001) & -0.194531 (<0.0001) & 0.00945 (<0.0001) & -0.000169 (<0.0001) \\
    & 1.187222 (<0.0001) & -0.085435 (<0.0001) & 0.001774 (<0.0001) & -0.166812 (<0.0001) & 0.007237 (<0.0001) & -0.000124 (<0.0001) \\
    & 1.386118 (<0.0001) & -0.104169 (<0.0001) & 0.00231 (<0.0001) & -0.162555 (<0.0001) & 0.006191 (<0.0001) & -9.3e-05 (0.003633) \\
    & 1.304145 (<0.0001) & -0.097004 (<0.0001) & 0.002137 (<0.0001) & -0.188486 (<0.0001) & 0.008582 (<0.0001) & -0.00015 (<0.0001) \\
    & 1.229122 (<0.0001) & -0.087063 (<0.0001) & 0.001764 (<0.0001) & -0.179439 (<0.0001) & 0.007615 (<0.0001) & -0.000125 (<0.0001) \\
    & 1.288569 (<0.0001) & -0.096806 (<0.0001) & 0.002168 (<0.0001) & -0.16137 (<0.0001) & 0.006719 (<0.0001) & -0.000114 (0.000385) \\
    & 1.138803 (<0.0001) & -0.077777 (<0.0001) & 0.001482 (<0.0001) & -0.182281 (<0.0001) & 0.007682 (<0.0001) & -0.000128 (<0.0001) \\
    & 1.355529 (<0.0001) & -0.102935 (<0.0001) & 0.00233 (<0.0001) & -0.195753 (<0.0001) & 0.009209 (<0.0001) & -0.000162 (<0.0001) \\
    & 1.286729 (<0.0001) & -0.095252 (<0.0001) & 0.002084 (<0.0001) & -0.173796 (<0.0001) & 0.00748 (<0.0001) & -0.000122 (0.000128) \\
    & 1.284461 (<0.0001) & -0.094719 (<0.0001) & 0.002058 (<0.0001) & -0.154055 (<0.0001) & 0.006413 (<0.0001) & -0.000106 (0.001001) \\
    & 1.356769 (<0.0001) & -0.100482 (<0.0001) & 0.002188 (<0.0001) & -0.176943 (<0.0001) & 0.007307 (<0.0001) & -0.000116 (0.000256) \\
    & 1.343944 (<0.0001) & -0.101608 (<0.0001) & 0.002291 (<0.0001) & -0.191069 (<0.0001) & 0.008108 (<0.0001) & -0.000134 (<0.0001) \\
    & 1.181608 (<0.0001) & -0.084683 (<0.0001) & 0.001756 (<0.0001) & -0.167864 (<0.0001) & 0.006736 (<0.0001) & -0.000107 (0.000866) \\
    & 1.279754 (<0.0001) & -0.09312 (<0.0001) & 0.001983 (<0.0001) & -0.200026 (<0.0001) & 0.009071 (<0.0001) & -0.000156 (<0.0001) \\
    & 1.336283 (<0.0001) & -0.100883 (<0.0001) & 0.002282 (<0.0001) & -0.159467 (<0.0001) & 0.005615 (0.000171) & -7.9e-05 (0.012905) \\
    & 1.231404 (<0.0001) & -0.089336 (<0.0001) & 0.0019 (<0.0001) & -0.234684 (<0.0001) & 0.012389 (<0.0001) & -0.000233 (<0.0001) \\
    & 1.259017 (<0.0001) & -0.090834 (<0.0001) & 0.001906 (<0.0001) & -0.16242 (<0.0001) & 0.006533 (<0.0001) & -0.000107 (0.000706) \\
    & 1.286087 (<0.0001) & -0.095839 (<0.0001) & 0.002116 (<0.0001) & -0.171082 (<0.0001) & 0.007834 (<0.0001) & -0.00014 (<0.0001) \\
    & 1.243217 (<0.0001) & -0.090548 (<0.0001) & 0.001924 (<0.0001) & -0.187799 (<0.0001) & 0.008103 (<0.0001) & -0.000132 (<0.0001) \\
    & 1.464282 (<0.0001) & -0.114637 (<0.0001) & 0.002709 (<0.0001) & -0.180623 (<0.0001) & 0.008349 (<0.0001) & -0.000147 (<0.0001) \\
    & 1.275468 (<0.0001) & -0.096105 (<0.0001) & 0.002154 (<0.0001) & -0.1627 (<0.0001) & 0.006911 (<0.0001) & -0.000116 (0.00027) \\
    & 1.316188 (<0.0001) & -0.099817 (<0.0001) & 0.002253 (<0.0001) & -0.177192 (<0.0001) & 0.007406 (<0.0001) & -0.000121 (0.000144) \\
    & 1.26229 (<0.0001) & -0.091236 (<0.0001) & 0.001942 (<0.0001) & -0.169022 (<0.0001) & 0.007051 (<0.0001) & -0.000119 (0.000196) \\
    & 1.205653 (<0.0001) & -0.086532 (<0.0001) & 0.001799 (<0.0001) & -0.211661 (<0.0001) & 0.009996 (<0.0001) & -0.000171 (<0.0001) \\
    & 1.235243 (<0.0001) & -0.090141 (<0.0001) & 0.001909 (<0.0001) & -0.165754 (<0.0001) & 0.006425 (<0.0001) & -0.0001 (0.001471) \\
    & 1.342381 (<0.0001) & -0.101417 (<0.0001) & 0.002301 (<0.0001) & -0.211247 (<0.0001) & 0.009795 (<0.0001) & -0.000165 (<0.0001) \\
    & 1.291467 (<0.0001) & -0.094846 (<0.0001) & 0.002044 (<0.0001) & -0.19375 (<0.0001) & 0.008863 (<0.0001) & -0.00015 (<0.0001) \\
    & 1.439101 (<0.0001) & -0.10984 (<0.0001) & 0.002512 (<0.0001) & -0.177249 (<0.0001) & 0.007251 (<0.0001) & -0.000116 (0.000272) \\
    & 1.306405 (<0.0001) & -0.098516 (<0.0001) & 0.002212 (<0.0001) & -0.131887 (<0.0001) & 0.004333 (0.004139) & -5.7e-05 (0.075286) \\
    & 1.196009 (<0.0001) & -0.085871 (<0.0001) & 0.001801 (<0.0001) & -0.166548 (<0.0001) & 0.00645 (<0.0001) & -9.9e-05 (0.002066) \\
    & 1.271511 (<0.0001) & -0.094543 (<0.0001) & 0.002093 (<0.0001) & -0.156238 (<0.0001) & 0.006067 (<0.0001) & -9.5e-05 (0.003111) \\
    & 1.204346 (<0.0001) & -0.0888 (<0.0001) & 0.001933 (<0.0001) & -0.162184 (<0.0001) & 0.00627 (<0.0001) & -9.9e-05 (0.001874) \\
    & 1.268871 (<0.0001) & -0.092455 (<0.0001) & 0.001986 (<0.0001) & -0.144969 (<0.0001) & 0.005069 (0.000789) & -6.8e-05 (0.034197) \\
    & 1.263455 (<0.0001) & -0.090325 (<0.0001) & 0.001875 (<0.0001) & -0.191276 (<0.0001) & 0.008352 (<0.0001) & -0.000137 (<0.0001) \\
    & 1.204791 (<0.0001) & -0.086846 (<0.0001) & 0.001811 (<0.0001) & -0.15736 (<0.0001) & 0.006807 (<0.0001) & -0.000122 (0.000102) \\
    \hline
    \multicolumn{7}{l}{*Note that the values for the original data slightly deviate from the values determined by Perrar \textit{et al.} \cite{perrar20donaldtrends}, as we re-built the polynomial mixed-effects models in R,} \\
    \multicolumn{7}{l}{that were originally coded in SAS.} \\
     \label{tab:raw_values_trend_analyses}
\end{longtable}
\end{small}
\end{landscape}

\begin{figure}[h!]
     \centering
     \begin{subfigure}[b]{0.49\textwidth}
         \centering
    \includegraphics[width=1\linewidth]{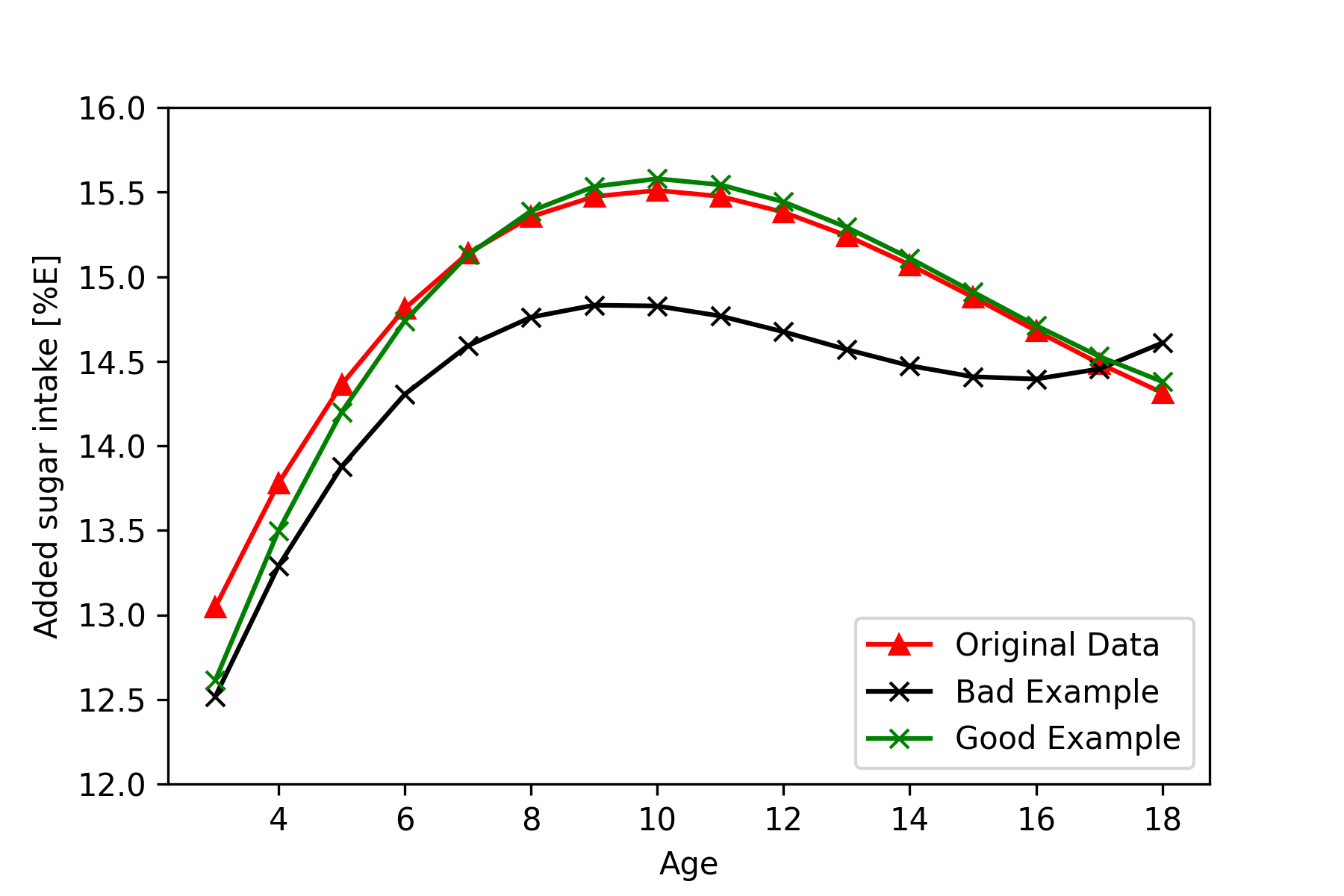}
    \caption{$N=1,312$}
    \label{fig:age_trend_as_cherry_1312}
    \end{subfigure}
    \begin{subfigure}[b]{0.49\textwidth}
         \centering
    \includegraphics[width=1\linewidth]{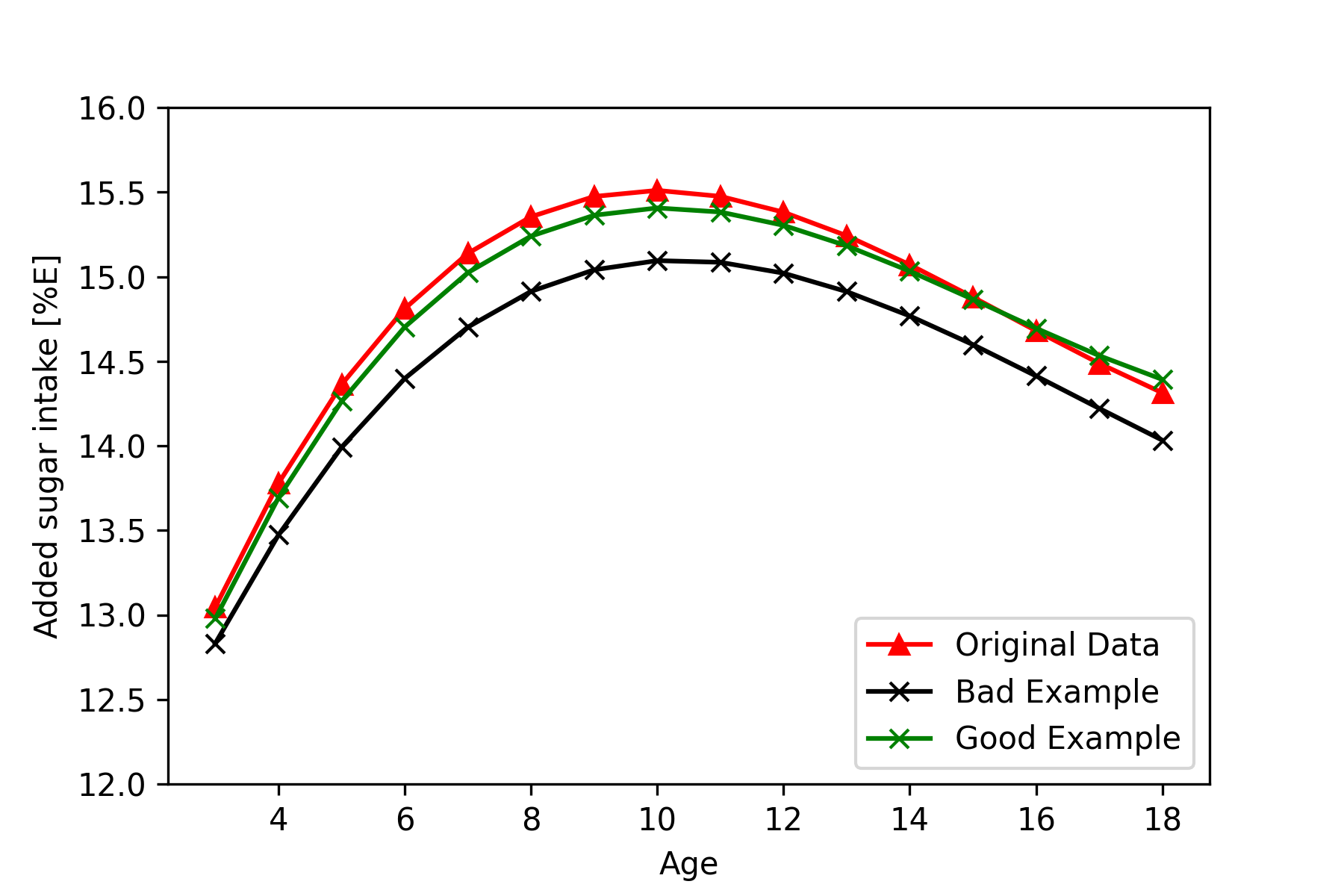}
    \caption{$N=10,000$}
    \label{fig:age_trend_as_cherry_10000}
    \end{subfigure}
    \caption{Effects of sample size for the stability of determined age trends. The good and bad examples (shown in green and black, respectively) are manually picked from a set of 100 samples for both sample sizes.}
    \label{fig:age_trends_as_cherrye}
\end{figure}

\begin{figure}[h!]
     \centering
     \begin{subfigure}[b]{0.49\textwidth}
         \centering
    \includegraphics[width=1\linewidth]{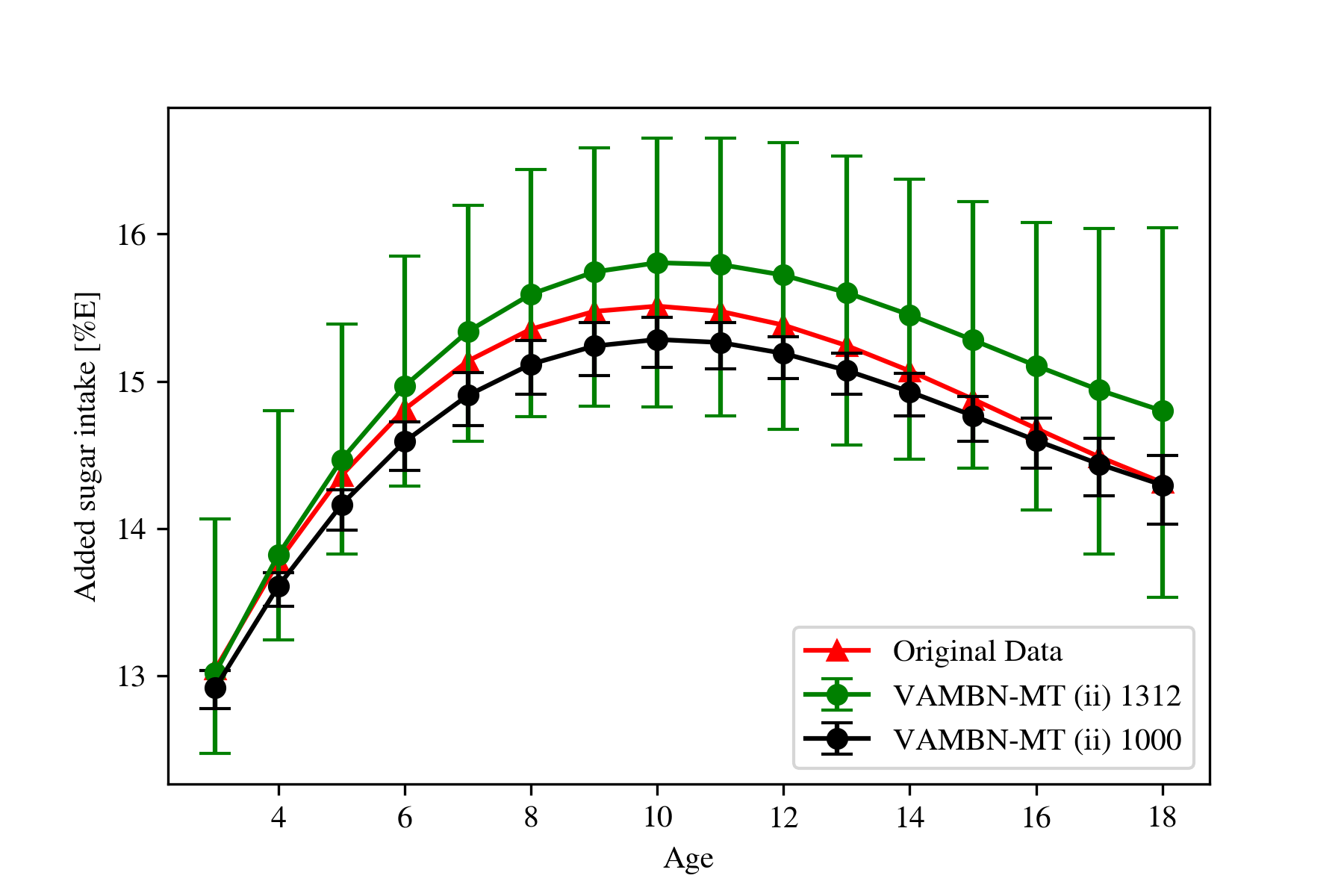}
    \caption{Age trend}
    \label{fig:age_trend_as_sample_size}
    \end{subfigure}
    \begin{subfigure}[b]{0.49\textwidth}
         \centering
    \includegraphics[width=1\linewidth]{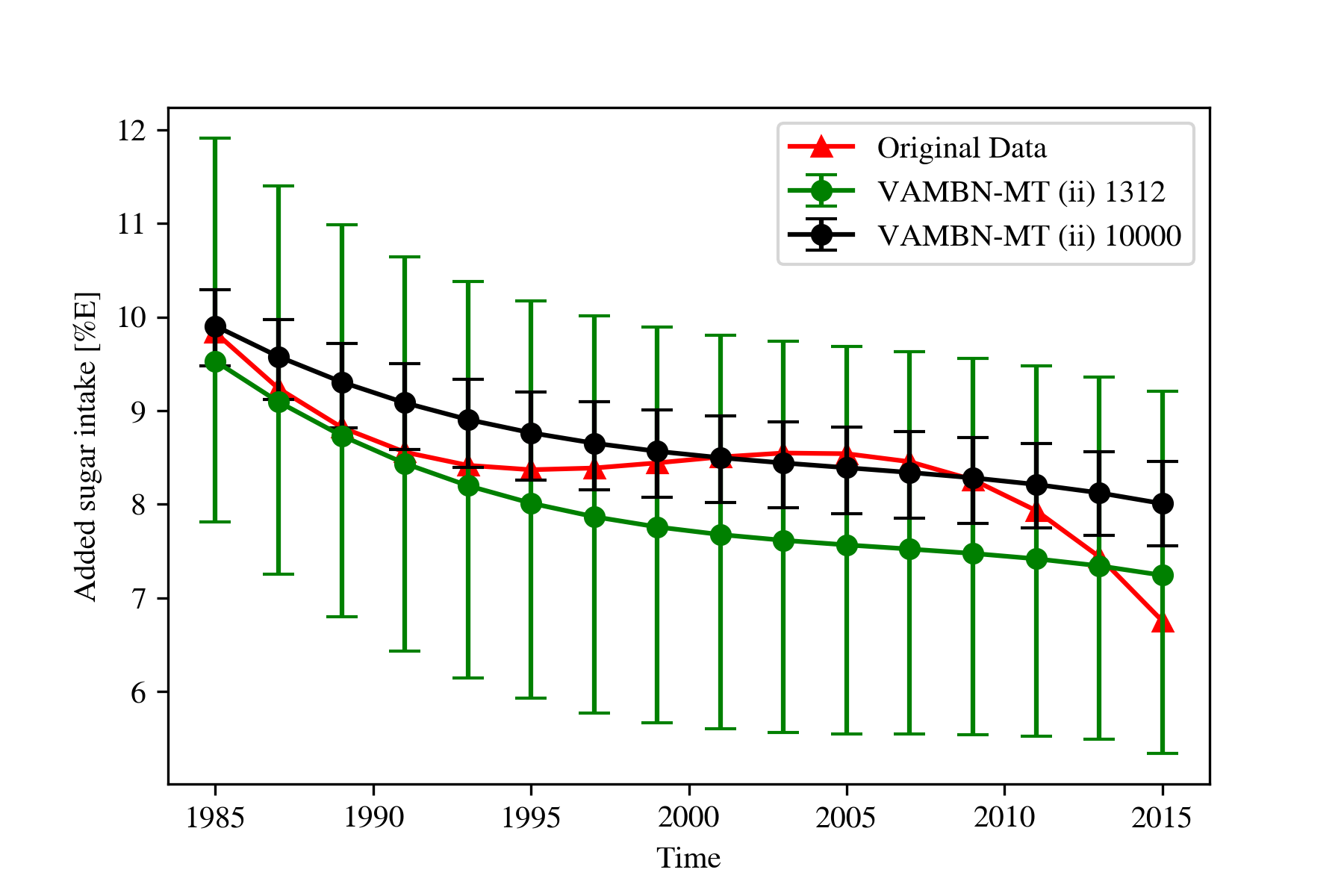}
    \caption{Time trend}
    \label{fig:time_trend_as_sample_size}
    \end{subfigure}
    \caption{Effects of sample size for age and time trends for added sugar intake predicted by polynomial mixed-effect regression models. Similar to Fig.~\ref{fig:trends_as}, we show averaged age and time trends of the synthetic datasets in comparison the the real data. Additionally, we show the results for two sample sizes, i.e., 1,312 (green) and 10,000 (black). For the smaller sample size (that corresponds to the original data size), also missingness is inserted.}
    \label{fig:trends_as_sample_size}
\end{figure}

\begin{figure}[h!]
     \centering
     \begin{subfigure}[b]{0.49\textwidth}
         \centering
    \includegraphics[width=1\linewidth]{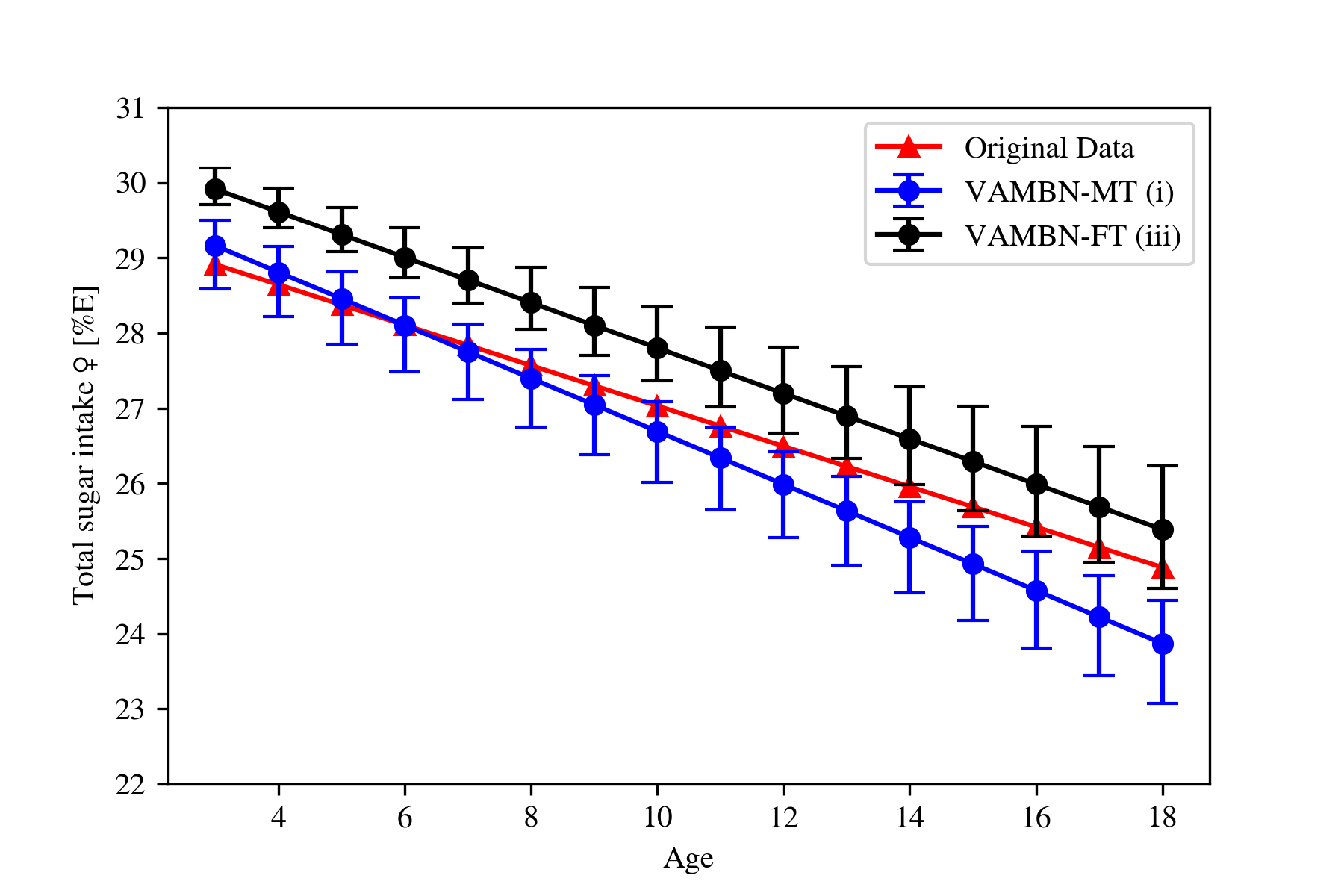}
    \caption{Age trend}
    \label{fig:age_trend_ts_boys}
    \end{subfigure}
    \begin{subfigure}[b]{0.49\textwidth}
         \centering
    \includegraphics[width=1\linewidth]{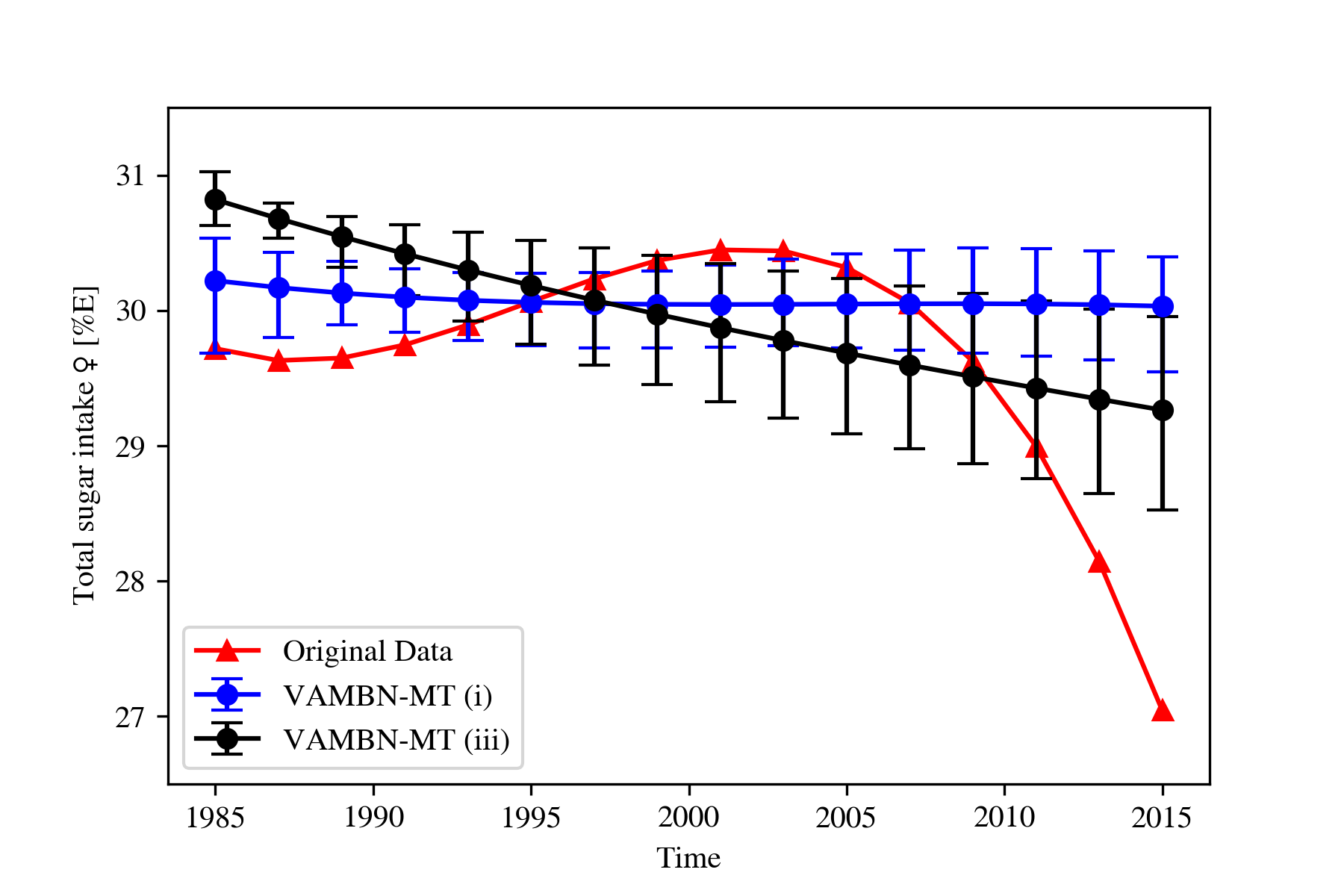}
    \caption{Time trend}
    \label{fig:time_trend_ts_boys}
    \end{subfigure}
    \caption{Age and time trends for the total sugar consumption. Whereas the synthetic data can reproduce the age trend, they fail to reproduce the time trend, independent on the module selection (i.e. whether time and total sugar are learned in one module or not). In the real data, only the age trend shows statistically significant values for all terms (linear, quadratic and cubic). For the time trend, both linear and quadratic show non-significant results.}
    \label{fig:trends_ts}
\end{figure}